%% file: imod.tex
\input o.tex

\longmodetrue
\refsin

\def\Ci{\CC^\infty } 
\def\Ca{\CC^\omega } \def\CaX{\Ca(X)}
\def\Germ{\CO_a}
\def\Powers{\C[\mskip-2.7mu[x^1- a^1,\ldots,x^m-a^m]\mskip-2.5mu]}

\def\M{\CM} \def\D{\CD} \def\E{\CE} \def\A{\CA} \def\K{\CK} \def\P{\CP} 
\def\F{\CF} \def\G{\CG} \def\Q{\CQ}

\def\q#1{q\ps{#1}} \def\qn{\q n}
\def\Fq#1{\F^{\times \q{#1}}} \def\Fqn{\Fq n}

   \def\EX{\E(X)}  \def\GX{\G(X)}
\def\Dn{\D^n} \def\An{\A^n}  

\def\PM{\P(\M)} \def\AM{\A(\M)} \def\AnM{\An(\M)}
\def\KM{\K(\M)} \def\NM{\CN(\M)} \def\QM{\Q(\M)} 
\def\dM{\M\Upstar }
\def\dMtilde{\widetilde{\M}\Upstar}

\def\v{\bo v}

\def\sf{\{f\}}

\def\W#1{\bo W_{#1}} \def\Wn{\W n}
\def\Wro{\W{r-1}}  \def\Wrox{\Wro(x)}
  
\def\Wr#1{\rho_{#1}} \def\Wrn{\Wr n} \def\Wrnx{\Wr n(x)}
\def\Wrg#1{\rho\Upstar _{#1}} \def\Wrgn{\Wrg n}
 
\def\Wo{\omega } \def\Wox{\Wo(x)} 
\def\Wvo{\varpi } \def\Wvox{\Wvo(x)} 
\def\WI{\W{\bo I}}  
\def\MW#1{\bo M_{#1}} \def\MI{\MW{\bo I}} \def\MIx{\MI(x)}
\def\vo#1{\varpi_{#1}} \def\voh{\vo h}  \def\vohx{\voh(x)}
\def\voI{\vo{\bo I}} \def\voIx{\voI(x)}

\def\g{\frak g}

\def\Hstar{\,*\>}

\def\dop{S}
\operator{End} \def\EndM{\End(\M)}

\newfam\msafam 
\font\tenmsa=msam10 \textfont\msafam=\tenmsa  

\edef\msahx{\hexnumber\msafam}
\mathchardef\Subset="3\msahx62
\vglue -.5in
\Title
Invariant Modules and the Reduction\\of Nonlinear Partial Differential Equations \\
to Dynamical Systems.

\author
Niky Kamran\\
Department of Mathematics\\
McGill University\\
Montr\'eal, Qu\'ebec \quad H3A 2K6\\
CANADA\\
\email nkamran@math.mcgill.ca
\support an NSERC Grant.

\author
Robert Milson\\
Department of Mathematics\\
Dalhousie University\\
Halifax, Nova Scotia \quad B3H 3J5\\
CANADA\\
\email milson@mscs.dal.ca
\support an NSERC Grant.

\pjoi

\printauthor
\dated
\vglue-.6in
\Abstract 
We completely characterize all nonlinear partial
differential equations leaving a given finite-dimensional vector space of
analytic functions invariant.  Existence of an invariant  subspace leads to a reduction of
the associated dynamical partial differential equations to a system of ordinary differential
equations, and provide a nonlinear counterpart to quasi-exactly solvable quantum Hamiltonians.
These results rely on a useful extension of the classical Wronskian determinant condition for linear
independence of functions. In addition, new approaches to the characterization of the annihilating
differential operators for spaces of analytic functions are presented.  

\page
\vglue .25in

\Section i Introduction.

The construction of explicit solutions to partial differential equations by symmetry
reduction dates back to the original work of Sophus Lie, \rf{Liepde}.  The reduction
of partial differential equations to ordinary differential equations was generalized
by Clarkson and Kruskal, \rf{CK}, in their direct method, which was later shown,
\rf{Odr}, to be included in the older Bluman and Cole nonclassical symmetry reduction
approach, \rf{BCp}. A survey of these methods as of 1992 can be found in \rf{OBirkhoff}.
Meanwhile, Galaktionov,
\rf{Galakis}, introduced the method of nonlinear separation that reduces a partial differential
equation to a system of ordinary differential equations, developed in further depth in
\rf{Galakis,GalakPos,GPS1,GPS2}.  Similar ideas appear in the work of King, \rf{King}, and the
``antireduction'' methods introduced by Fushchych and Zhdanov,
\rf{FushZh,Fush95}. Svirshchevskii,  \rf{Sv,Svx,Svi}, made the
important observation that one could, in the one-dimensional case, characterize in terms of  higher order
(or generalized) symmetries those nonlinear ordinary differential operators  that admit a given invariant
subspace leading to nonlinear separation.

In quantum mechanics, linear differential operators with invariant subspaces
form the foundation of the theory of quasi-exactly solvable (QES) quantum models as
initiated by Turbiner, Shifman, Ushveridze, and collaborators,
\rf{Shifman,ShifTurb,Turbiner,U}.  The basic idea is that a Hamiltonian
operator which leaves a finite-dimensional subspace of functions invariant can be restricted to
this subspace, resulting in an eigenvalue problem which can be solved by
linear algebraic techniques.  The Lie algebraic approach to quasi-exactly
solvable problems requires that the subspace in question be invariant under a
Lie algebra of differential operators $\g$, in which case the Hamiltonian
belongs to the universal enveloping algebra of $\g$; see
\rf{GKO,GKOrqes,Oqes,TurbCRC,U} for details and \rf{IL,Levine} for applications to
molecular spectroscopy, nuclear physics, and so on.  
Zhdanov, \rf{Zhdanov}, indicated how one could characterize quasi-exactly solvable operators using
higher order symmetry methods.  We should also mention   Hel--Or and Teo,
\rf{HelTeo}, who have applied group-invariant subspaces in computer vision, naming
their elements ``steerable functions''.

Motivated by a problem of Bochner, \rf{Bochner}, to characterize differential operators having
orthogonal polynomial solutions, Turbiner, \rf{Turbpoly}, initiated the study of differential
operators leaving a polynomial subspace invariant. In one-dimension, the remarkable result is that
the operators leaving the entire subspace of degree $\leq n$ polynomials invariant are the
quasi-exactly solvable operators constructed by Lie algebra methods.  These results were further
developed for multidimensional and matrix differential operators, and difference operators by
Turbiner, Post and van den Hijligenberg, \rf{PostTurb,PostHijl,Turbpolz,Turbis}, and Finkel and 
Kamran, \rf{FKp}.

In this paper, we broaden the general theory of nonlinear separation to include partial differential
operators, and argue that it constitutes the proper nonlinear generalization of quasi-exactly
solvable linear operators.  Our theory provides an explicit characterization of
all nonlinear differential operators that leave a given subspace of functions
invariant. In the time-independent case, solutions lying in the subspace are obtained by
solving a system of nonlinear algebraic equations.  For evolution
equations and certain other dynamical partial differential equations, we show how explicit solutions in
the given subspace are obtained by reducing the partial differential equations to a finite-dimensional
dynamical system.  We illustrate our method with a number of significant examples.

Our methods can also be compared and contrasted with the more algebraic
theory of $D$-modules, \crf{BjorkD,Coutinho,Oaku,OakuTak}.  In
particular, we describe new algorithms for determining the annihilator
of a given finite-dimensional subspace, based on the study of
generalized Wronskian matrices and their ranks.  These methods derive
their justification from the general theory of prolonged group
transformations developed in \rf{E,Osing}.

The first section of the paper outlines the basic setting of our methods --- finite-dimensional
spaces of analytic functions defined on an open subset of real Euclidean space.  The case of analytic
functions of several complex variables is more subtle, and requires cohomological or geometric
restrictions on the domain. Section 3 presents the basic tools in our study: a multi-dimensional
generalization of the classical Wronskian condition for linearly independence of functions.  In fact,
these results form a very particular case of a general study of orbit dimensions of prolonged group
actions formulated in
\rf{Osing}.  The key concept is the notion of a ``regular'' space of analytic functions, and only for
such subspaces is one able to characterize the differential operators that annihilate the subspace or
leave it invariant.  While the Hilbert basis theorem is not generally applicable to ideals of
analytic differential operators, one can nevertheless algorithmically determine a finite generating set
of annihilating differential operators when the subspace is regular.   Section 4 contains our basic
approach to the annihilator, which relies on a useful extension which we
name an ``affine annihilator'', which is a differential operator that maps every function in the
subspace to a constant.  The construction ultimately rests on an interesting lemma characterizing
analytic solutions to a system of variable coefficient linear equations of constant rank; it is at
this critical point that the distinction between the real and complex analytic situations becomes
evident.  We include several examples illustrating our construction, including a convenient
characterization of the affine annihilators of simplicial subspace of monomials.  Section 5
applies these constructions to characterize all linear and nonlinear differential operators that
leave a given regular subspace invariant. In the scalar case,  Svirshchevskii, \rf{Sv,Svx,Svi},
characterized these differential operators using generalized symmetries, and we prove that
Svirshchevskii's symmetry operators coincide with our affine annihilators, thereby establishing the
generalization of Svirshchevskii's methods to analytic functions of several variables. 
Finally, section 6 outlines how our results can be applied to the construction of explicit solutions
to linear and nonlinear partial differential equations based on the method of nonlinear separation.

\Section m Function Spaces.

Let $X\subset\Rm$ be an open, connected subset of Euclidean space, with coordinates $\psups xm$. Our
basic set of allowable functions will the space $\F =\CaX$ of analytic real-valued functions
$f\colon X\to \R$. We may regard $\F$, depending on the
circumstances, as either a real vector space, or as an algebra over
the reals.

Even though we shall primarily restrict our attention to real domains and real
analytic functions, much of the exposition can be adapted to other
situations. For instance, a much easier situation is when
$\F = \Germ$ is the space of all germs of analytic functions at a single point $a = \psups am \in
X$, or, equivalently, the space $\Powers$ of convergent power series at the point $a$.  Another
interesting example, studied extensively in $D$-module theory, is when $\F$ is the field
of meromorphic functions on $X$, or, more generally, any differential field, \crf{Kolchin}.  The
case when $\F = \Ci(X)$ consists of smooth functions on $X$ is also quite interesting, but much
more difficult to treat owing to a number of pathologies that do not appear in the analytic
context.  With the proper restrictions, our methods and results can also be made to apply to
various function spaces in the complex-analytic category.  We briefly indicate how this may be done
in Section 4, below.

Once we have fixed the function space $\F$, our primary object of
study are finite-dimensional subspaces $\M\subset \F$. In particular, given functions $\subs fk
\in \F$, we let $\M = \bsubs fk$ denote the subspace spanned thereby. We shall use
$r$ to denote the dimension of $\M$, and often use $\subs fr \in \F$ to
denote a basis.  In applications to quasi-exactly solvable quantum
problems, $\M$ is a finite dimensional module for the action of a
transformation group $G$ on $X$, \crf{GKO,GKOrqes}.

Let $\D = \D(\F)$ denote the space of linear differential operators
whose coefficients belong to the function space $\F$.  The
multiplication of differential operators is defined in the usual
manner, making $\D$ into an associative, but noncommutative algebra
over $\F$ --- the latter acting by left multiplication. Furthermore,
the operators in $\D$ define linear maps $L\colon \F \to \F$, and so
$\F$ will also be regarded as a $\D$-module. There is a natural
filtration on $\D$, with $\Dn$ denoting the $\F$-submodule consisting
of differential operators of order $\leq n$.  The domains $X$
considered here are such that every linear differential operator $L
\in \Dn$ is realized as a finite sum
$$L = \sum _I h_I(x)\> \partial _I, \where 
\partial _I = {\partial ^k \over \partial{x_{i_1}}\cdots \partial{x_{i_k}}},
\Eq{dop}$$ 
and the coefficients $h_I\in \F$.  The sum in \eq{dop} is over symmetric multi-indices $I = \psubs
ik$ of orders
$0\leq k =\#I \leq n$.  Since $\D^{n+1} = \D^1 \cdot \Dn$, we may identify
the quotient $\Dn/\D^{n-1}$ with the $\F$-module of homogeneous differential operators
of order $n$.

We note that there are $q_n = {m + n - 1 \choose n}$ different
symmetric multi-indices $I$ of order $\#I = n$, which is the number of
different \nth order partial derivatives $\partial _I$.  Similarly,
there are 
$$\qn = q_0 + q_1 + \ldots + q_n = {m + n \choose n}\Eq{qn}$$
different symmetric multi-indices $I$ of order $\#I \leq n$.  An \nth
order differential operator is uniquely determined by its $\qn$
different coefficients $h_I(x)$, $\#I\leq n$.  Therefore, we can
identify $\Dn$ with the space $\Fqn$ --- the Cartesian product of
$\qn$ copies of $\F$.  Explicitly, the isomorphism $\sigma _n\colon
\Dn \simarrow \Fqn$ maps a linear differential operator \eq{dop}
to the column vector
$$\sigma _n(L) = \bo h(x) = (\;\ldots, h_I(x), \ldots\;)^T \Eq{sig}$$
whose entries, indexed by the multi-indices of order $\#I\leq n$, are
the coefficients of $L$. 


\Section W Wronskians and Stabilization.

The most basic necessary and sufficient condition for the linear independence of
solutions to a homogeneous linear scalar ordinary differential equation is the nonvanishing of
their Wronskian determinant, \crf{Hale}. Many of our results will rely on a significant
multi-dimensional generalization of this classical Wronskian lemma, which can be applied to
\iz{any} collection of analytic functions.  We refer to
\rf{Osing} for details on the following definitions and results, including extensions to both
smooth and vector-valued functions.

\Df{Wronskian} The \nth order \is{Wronskian matrix} of the functions
$\subs fr \in \F$, is the $r \times \qn$
matrix
$$\Wn(x) = \Wn\csubs fr(x) = \pmatrix{ f_1(x)&\ldots&\partial_I
f_1(x)&\ldots&\ \cr \vdots&\ddots&\vdots&\ddots&\cr
f_r(x)&\ldots&\partial_I f_r(x)&\ldots\cr },\Eq{Wmatrix}$$ whose
entries consist of the partial derivatives of the $f_\kappa$'s
with respect to the $x^i$'s of all orders $0 \leq \#I \leq n$.

In the scalar case $X\subset \R$,  the standard Wronskian determinant 
coincides with the determinant of the \xsts{r-1}order Wronskian matrix
$\Wro = \Wro\csubs fr$, which happens to be a square matrix.   

To this end, we define the \is{Wronskian matrix rank} function
$$\Wrnx = \rank \Wn\csubs fr(x).  \Eq{Wrank}$$
Note that $\Wrnx$ only depends only on the subspace $\M = \bsubs fr$ spanned by the given functions,
and not on the particular generators or basis. In particular, $0 \leq \Wrnx \leq \dim \M$. 
Moreover,
$\rho _n(x)$ is lower semi-continuous: if $\Wrn(x_0) = k$, then $\Wrnx
\geq k$ for all $x$ in a sufficiently small neighborhood of $x_0$.
The \is{generic Wronskian rank} of order $n$ is
$$\Wrgn = \max \set{\Wrnx}{x\in X}.$$
The first key result is the following:

\Th{Wr} Let $\M\subset \F$ be an $r$-dimensional subspace and $\Wrgn$, $n\geq 0$, be the sequence of its
generic Wronskian ranks.  Then $\Wrgn = r = \dim \M$ for $n \gg 0$ sufficiently large.  Moreover, if we
define the (generic) \is{stabilization order} $s= \min \set n{\Wrgn = r}$ of  $\M$ to be the
minimal such $n$, then we find
$$\Wrg0 < \Wrg1 < \cdots < \Wrg{s-1} < \Wrg s = \Wrg{s+1} = \Wrg{s+2} =
\cdots = r = \dim \M. \Eq{rs}$$

\smallskip
In other words, once the generic Wronskian ranks become equal,
they stabilize, and achieve a value equal to the dimension of the subspace.
 In particular, the ranks cannot ``pseudostabilize'', \rf{E,Osing}, and so 
\iz{must} strictly increase before stabilization sets in.
 It is entirely possible, though, that they increase only by $1$
at each order --- an evident example occurs when all the functions only depend on
a single variable.

While knowledge of the stabilization order reduces the amount of work
required to compute the order of the generic Wronskian rank, one can,
in all cases, replace $s$ by $r-1$.  Thus, one has:

\Co{Wr1} The analytic functions $f_1(x),\ldots ,f_r(x)$ are linearly
independent if and only if the generic rank of their \xsts{r-1}order
Wronskian matrix $\Wrox = \Wro\csubs fr(x)$ is equal to $r$.

\Proof Indeed, \lm{Wr} implies that $\subs fr$ are linearly
independent if and only if $\Wrgn = r$ for any $n\geq s$ greater than
or equal to the stabilization order.  Moreover, \eq{rs} implies that
the stabilization order of an $r$-dimensional subspace $\M\subset \F$
is always bounded by $s\leq r-1$. Thus, one typically does not need to
compute the rank of the order $n = r-1$ Wronskian to detect linear
independence --- checking at the stabilization order is sufficient.
Another way of stating this result is the following: if the generic
ranks of the Wronskians $\W k(x)$ and $\W{k+1}(x)$ are equal, then
this rank is the same as the dimension of $\M$; moreover, this
equality first occurs when $k = s\leq r-1$.\qed

Kolchin, \rf{Kolchin; p. 86}, states and proves a version of \co{Wr1}
assuming that the functions belong to a differential field, \eg the
field of meromorphic functions.  It would be of interest to adapt our
constructions to this case.  

It is noteworthy that \th{Wr} is a particular case of a general theorem governing the orbit
geometry of prolonged transformation groups acting on jet bundles, described in detail in \rf E. 
In the present situation, the relevant group is the elementary
$r$-parameter abelian group 
$$(x,u) \longmapsto \left(x,u+\sum_{\kappa =1}^r\; t_\kappa f_\kappa(x)\right),\Eq{atg}$$
acting on $E = X\times \R$.
The dimension of the prolonged group orbits contained in the \nth order jet fiber $\Jn
E\at x$ equals the Wronskian rank $\Wrnx$.  Moreover, in the analytic category, the group
\eq{atg} acts effectively
if and only if the subspace
$\M \bsubs fr$ has dimension $r$. The extensions of this result to more general smooth group actions
are also treated in \rf{Osing}, and can be applied to the more subtle case of subspaces $\M
\subset \Ci$ containing smooth functions.  For brevity, we shall refrain from discussing this
more complicated case here.


Although, for an $r$-dimensional subspace $\M$, the \xsts{r-1}order
Wronskian has generic rank $r$, it may certainly have lower rank at particular
points in the domain $X$.  A simple example is provided by the
functions $f_1(x) = 1$, $f_2(x) = e^{x^2}$, which have first order
Wronskian determinant $\det \W1(x) = 2x e^{x^2}$, which is singular at
$x=0$.  (The classical Wronskian lemma implies that these functions cannot be common solutions to a
regular, homogeneous, second order linear differential equation.)  Our
applications typically require that, for some $n$ sufficiently large,
the Wronskian rank $\rho _n(x)$ be equal to $r$ at every $x\in X$, and so we
need to determine when this occurs. In the preceding example,
$$\W2[f_1,f_2](x) = \pmatrix{f_1&f_1'&f_1''\cr f_2&f_2'&f_2''\cr} = 
\pmatrix{1&0&0\cr e^{x^2}&2x e^{x^2}&(2 + 4x^2) e^{x^2}\cr}$$
 has rank $2$ everywhere, as do all the higher order Wronskian matrices.
Therefore, it is of interest in understanding, not just how the generic Wronskian ranks behave,
but also how the
$\Wrnx$ behave at a single point. 

\Th{Wo} If $\dim \M = r$ and $x\in X$ is any point, then there exists a \is{finite} integer $n =
n(x)$ such that $\Wr k(x) = r$ for all $k\geq n$.  We call $n$ the \is{Wronskian order} of the point
$x$, and denote it by
$$\Wox = \min \set n{\Wrnx = r}. \Eq{wo}$$
\smallskip

For example, if we take $f_1(x) = 1$, $f_2(x) = e^{x^k}$, for $k\geq
2$ an integer, then the Wronskian order of $x=0$ is equal to
$\Wo(0) = k$.  Therefore, the Wronskian order of a point can be
arbitrarily large.

Again, \th{Wo} is a corollary of a more general theorem about the
geometry of prolonged transformation groups.  In the language of
\rf{Osing}, a point at which $\Wr k(x) < r$ for \iz{all} $k$ would be
known as a ``totally singular point'' for the associated
transformation group \eq{atg}.  However, Theorem 6.4 of \rf{Osing}
states that an analytic transformation group admits no totally
singular points.  In fact, for the elementary group \eq{atg}, this result is not hard to prove
directly --- it is a consequence of the basic result that a nonzero analytic function cannot
have all zero derivatives at a single point.  And this is the main reason
why we must restrict out attention to analytic, as opposed to smooth,
functions.

Given an $r$-dimensional subspace $\M = \bsubs fr$, \co{Wr1} implies that most
 --- meaning those belonging to a dense open subset of $X$ --- points have
Wronskian order $r-1$ or less.\fnote{If $\dim X = 1$, then the
Wronskian order is never less than $r-1$.  More generally, if $\dim X
= m$, then the minimal Wronskian order $l$ of any subspace is bounded
from below by the inequality $q\ps l = {m+l\choose l}\geq r$.}  The exceptions
are the points $x$ where $\Wrox$ has less than maximal rank. This permits us to formulate a basic
estimate on the Wronskian order of the space $\M $. 
In general, given an ordered $r$-tuple $\bo I = \psubs Ir$ consisting of symmetric
multi-indices, we define its order to be $\#\bo I = \max \set{\#I_\nu}{\nu=1\ldots
r}$.  Let $\WI(x)$ denote the associated $r\times r$ submatrix of $\W k(x)$
whose columns are indexed by the multi-indices $I_\nu $ in $\bo I$.  More
prosaically, the entries of $\WI(x)$ are the partial derivatives
$\partial _{I_\nu } f_\mu $, $\mu ,\nu =1,\ldots ,r$, indicated by
$\bo I$. Note that $\WI(x)$ depends on at most $\#\bo I$\upth order derivatives of the functions
$\subs fr$.  The \xth{\bo I}\is{Wronskian minor} is then
$$\MI(x) = \det \WI(x). \Eq{MI}$$
  Clearly, the Wronskian order of a point $x$ is the smallest
multi-index order with nonvanishing Wronskian minor, \ie
$$\Wox = \min \set{\#\bo I}{\MI(x) \ne 0}. \Eq{Wox}$$

Given a non-zero analytic function $h(x)\nequiv 0$, we define its
\is{vanishing order} $\vohx$ at a point to be the order of its first
nonzero terms in the power series expansion at $x$.  Alternatively, we
can define the vanishing order by differentiation:
$$\vohx = \min \set{\#I}{\partial _Ih(x) \ne  0}. \Eq{vox}$$
In particular, $\vohx = 0$ if $h(x) \ne  0$.  
Note that the vanishing order of a nonzero analytic function is always finite.

\Df{Wvo} Given a subspace $\M$, let $\voIx = \vo{\MI}(x)$ denote the vanishing order of its
\xth{\bo I}Wronskian minor $\MIx$.  The \is{Wronskian vanishing order} of $\M$ is defined as the
minimal vanishing order of the minors of $\Wrox$, so
$$\Wvox = \min \set{\voIx}{\#\bo I\leq r-1}. \Eq{Wvox}$$

\medskip
The Wronskian vanishing order of a subspace provides an immediate bound on its Wronskian order at
a point.

\Pr{WoWvo} Let $\dim \M = r$ and let $x\in X$.  If the Wronskian
vanishing order of $\M$ at $X$ is $\Wvox$, then the Wronskian order of
$x$ is bounded by
$$\Wox \leq r-1 + \Wvox. \Eq{WoWvo}$$

\Proof It suffices to note that if $J$ is any multi-index of order
$\#J = j$, then the \xth{J} order derivative of a Wronskian minor
$\MIx$ of order $\#\bo I = i$ can be written as a finite linear combination of  Wronskian minors
$$\partial _J\MIx = \sum _{\bo K}\ \MW{\bo K}(x) $$
of orders $\# \bo K\leq i+j$.  Hence, if $J$ and $\bo I$ are such that
$\partial_J\MIx\neq 0$, then there must exist a $\bo K$ with  $\# \bo K\leq i+j$ such that
$\MW{\bo K}(x)\neq 0$. \qed

\Ex{cos2x} Consider the three-dimensional subspace $\M = \{1,\cos
x,\cos 2x\}$.  The Wronskian determinant of these three functions is
$\det\W2(x) = -4 (\sin x)^3$, and hence $\M$ has Wronskian vanishing
order $3$ at $x = n\pi $, $n\in \Z$. The fourth order Wronskian matrix
$\W4(x)$ is found to have rank $3$ for all $x\in \R$, and hence the
Wronskian order at the singular points $x = n\pi $ is $4$.

If $x_0$ has Wronskian order $n$, then, by continuity, $\Wrnx = r$ for
all $x$ in some neighborhood of $x_0$.  However, a
global bound on the Wronskian order may not exist.  We therefore introduce the
following important definition.

\Df{reg} A subspace $\M$ is called \is{regular} if it has uniformly
bounded Wronskian order at each point, \ie there exists a finite $n$
such that $\Wrnx = r$ for \is{all} $x\in X$.  The minimal such $n$ is
called the \is{order} of $\M$.

Many of the basic results in this paper require the underlying regularity of the subspace.

\Ex{nonreg} Not every subspace is regular.  For example, let $X = \R$.  Consider the
one-dim\-en\-sion\-al subspace spanned by an analytic function
$f(x)$ which has a zero of order $k$ at $x_k$, for $k =
1,2,3,\ldots\ $, where $x_k \to \infty$ as $k \to \infty$; such a
function can be constructed using a Weierstrass product expansion,
\crf{Ahlfors; p. 194}.  Then $\rank \Wn[f](x_k) = 0$ for $n < k$,
while $\rank \Wn[f](x) = 1$ for $n \geq k$ and $x$ in any
neighborhood of $x_k$ that does not contain $x_{k+1}, x_{k+2}, \ldots\
$.  In other words, the Wronskian order of each $x_k$ is equal to $k$.

We know two straightforward mechanisms for proving that a given subspace is regular.
The key is to avoid an infinite sequence of points whose common
Wronskian order is unbounded, as in the preceding example.  The first
method relies \pr{WoWvo} for proving regularity.

\Pr{Wvreg} If $\M$ is a subspace with uniformly bounded Wronskian vanishing order, so $\Wvox \leq 
n$ for all $x\in X$, then $X$ is regular.

Alternatively, one can generate regular subspaces by restricting the domain of the functions and
using the basic properties of compactness.

\Pr{compact} If $\M \subset \CaX$ is any subspace, and $Y\Subset X$ is an open subset with
compact closure in $X$, then the subspace $\;\widehat {\M} = \M \mid _Y\>\subset \Ca(Y)$ obtained by
restriction to $Y$ is a regular subspace.

\Section{ann} Annihilators.

Our characterization of differential operators preserving spaces
of analytic functions relies on several intermediate constructions of
independent interest.  The first order of business is to characterize the
differential operators that annihilate all the elements of our subspace $\M$. 
We begin with the linear annihilating operators; their nonlinear counterparts will be treated
in the following section.

\Df{ann} The \is{annihilator} $\A = \AM$ of a subspace $\M\subset \F$
is the set of all linear differential operators that annihilate every
function in $\M$, so
$$\AM = \set{K\in \D}{K[f] = 0\ \roh{ for all }\ f\in \M}.$$
\smallskip

We will refer to the elements of $\AM$ as \is{annihilating operators}. 
We note that $\A\subset \D$ is, in fact, a left ideal, since if $K[f]
= 0$ and $L\in\D$ is any linear differential operator, then clearly $L[K[f]]= 0$ and so 
$L\cdot K \in\A$.

\Ex{AR}  Let $X \subset\R$.  Let $\M\subset \Ca(\R)$ be an $r$-dimensional  subspace. A classical
construction, \crf{Wil}, produces an \rth order annihilating operator
$$K_r = h_r(x) \partial ^r + h_{r-1}(x)\partial ^{r-1} + \cdots  + h_0(x).\Eq{Kr}$$
Indeed, introducing a basis $\subs fr$ of $\M$, then the conditions $K_r[f_\nu ] = 0$, $\rg \nu
r$, forms a homogeneous system of $r$ linear equations for the $r+1$ coefficients of $K_r$. 
Cramer's rule produces the solution
$$h_k(x) = (-1)^{r-k}\>\MW k(x), \Eq{hkx}$$
where $\MW k(x) = \det\W{01\ldots k-1,k+1\ldots r}(x)$ denotes the $r\times r$ Wronskian minor
obtained by deleting the \kth column of the $r\times (r+1)$ Wronskian matrix $\W r(x)$.
In particular, the leading term $h_r(x) = \MW r(x) = \det\Wrox$ is the classical Wronskian
determinant, and hence $K_r$ is a nonsingular differential operator if and only if the classical
Wronskian never vanishes, which implies that $\M$ is regular of the minimal possible order $r-1$.
\pari
If $h_r(x) \ne 0$ is never zero, then the
annihilator $\A$ is generated by $K_r$.  Indeed, to prove that every other
annihilating operator has the form $K = L\cdot K_r$, we note that
every linear ordinary differential operator can be written (uniquely)
in the form $T = Q\cdot K_r + R$, where $Q\in \D$ and $R \in \D^{r-1}$
has order at most $r-1$. (If $T$ has order $\leq r-1$, then $Q=0$.)
But $T\in \AM$ if and only if $R \in \AM$ is an annihilating operator.
But the dimension of the kernel of a linear
differential operator of order $k$ is at most\fnote{Operators with degenerate symbols may not
admit enough analytic solutions to span a full $k$-dimensional kernel.  For
example, the functions annihilated by the first order operator $L = x\partial _x + 1$ are multiples of
the non-analytic function $1/x$, and so $L$ has a zero dimensional (analytic) kernel.}  $k$, and so $R =
0$.
\pari
On the other hand, if $h_r(x)$ vanishes at points $x\in X$, then $K_r$ does \is{not}
provide a basis for the annihilator of $\M$.  For example, the
annihilator of the one-dim\-en\-sion\-al subspace spanned by the
function $f_1(x) = x$ is generated by the two differential operators
$x\partial -1$ and $\partial ^2$. (Although $\partial ^2 = \f x \partial \cdot (x\partial -1)$, the
operator $\f x \partial$ is not analytic, and so not allowed.) More complicated cases are discussed
below. 

\Ex{A1} Let $X\subset \Rx m$ and consider a one-dimensional subspace
$\M = \sf$ where $f\nequiv 0$.
The linear operators $K_i = f\partial _i - f_i$, $\rg im$, where $f_i
= \partial _if = \pd f{x_i}$, clearly belong to $\A^1$.  If $f(x) \ne
0$ never vanishes, then $\subs Km$ form a basis for $\A$.  The proof
of this fact is similar to the ordinary differential operator result
of \ex{AR}.  We first note that any differential operator can be
written in the form $T = \sum Q_i \cdot K_i + g$, where $\subs Qm \in
\D$ are differential operators, and $g \in \D^0 \simeq \Ca$ is a
multiplication operator.  Clearly $T\in \AM$ if and only if $g \equiv
0$, proving the result.  The case when $f$ vanishes on a subvariety of
$X$ is more subtle --- see below. Extensions to higher dimensional
subspaces are also discussed below.

  The filtration of $\D$ induces a filtration of the annihilator, and
we let $\An =\A\cap\Dn$ denote the subspace of annihilating operators
of order at most $n$.

\Pr{sigA} Let the analytic functions $f_1,\ldots ,f_r$ span a finite-dimensional
subspace $\M = \bsubs fr \subset\F$.  Let $\Wn$ the associated $\nth$ order
Wronskian matrix.  The map $\sigma _n$ defined by \eq{sig} defines an
isomorphism
$$\sigma _n\colon \ \An \simarrow \ker \Wn\,, \Eq{AWn}$$
between the \nth order annihilator of $\M$ and the common kernel 
$$\ker\Wn = \set{\bo h \in \Fqn}{\Wn\cdot \bo h = 0}.$$

\Proof It is sufficient to note that for every $K\in \AnM$ and
corresponding vector $\bo h = \sigma _n(K)$ with analytic entries, the \xth\kappa
entry of the matrix product $\Wn\cdot \bo h$ is equal to
$K[f_\kappa]$.  \qed

Had we taken $\F$ to be the algebra of power series (or analytic germs), then a
straightforward adaptation of the classical Hilbert basis theorem,
\rf{Hilbert,ZS} would prove that the annihilator $\AM$ of every finite-dimensional subspace
$\M\subset \F$ is finitely generated.  This is because the power series algebra is Noetherian,
\crf{ZS}. However, the algebra $\F = \CaX$ of globally defined analytic
functions is \iz{not} Noetherian, and so the Hilbert basis theorem
does not apply.  The following example shows that not every ideal of  $\F = \CaX$ is finitely
generated.

\Ex{nonfg} Consider again the analytic function $f(x)$ introduced in \ex{nonreg}.  Let $f^{(k)}$,
$k=0,1,2,\ldots$ denote the corresponding derivatives and note that
$f^{(k)}\partial - f^{(k+1)}$ is an annihilating operator for all $k$.
It isn't hard to see that the ideal of $\F = \CaX$ that is generated by all
the $f^{(k)}$ is not finitely generated, and therefore, in this case, the annihilator
$\A$ of $\M = \{f\}$ is not a finitely generated ideal of $\D$.

Later we will prove that the annihilator ideal of a \is{regular} subspace is finitely
generated.  to this end, we  introduce a useful extension of the notion of the annihilator.

\Df{can} The \is{affine annihilator} $\KM\subset \D$ of a subspace $\M\subset
\F$ is the subspace of those linear differential operators that
map every function in $\M$ to a constant function:
$$\KM = \set{L\in \D}{L[f] = c \in \R,\ \roh{ for all }\ f\in \M}.$$

Note that, as defined, the affine annihilator $\KM$ is a vector space,
rather than an $\F$-module.  Considered as a vector space, the annihilator $\AM$  is evidently a
subspace of
$\KM$.  The difference between the two subspaces has a natural interpretation.

\Df{odual} The \is{operator dual} to the subspace $\M$ is the quotient
vector space
$$\dM = \KM/\AM. \Eq{odual}$$

Since we are quotienting by the annihilator, there is a natural action
of $\dM$ on $\M$ itself induced by the action of $\KM$.  In this
fashion, if $\M$ is finite-dimensional, then there is a natural linear
injection from $\dM$ into the abstract dual of $\M$.  As the following
theorem will demonstrate, regularity implies that this injection is,
in fact, an isomorphism.

\Th{dM} If $\M\subset\F$ is a regular $r$-dimensional subspace of order
$s$, then its operator dual $\dM$ is also $r$-dimensional.  Moreover, if
$\subs fr$ forms a basis of $\M$, then there exists a dual basis for
$\dM$ represented by differential operators $\subs Lr$ of order at
most $s$ such that
$$L_i(f_j) = \delta ^i_j,\qquad \rg {i,j}r. \Eq{dualops}$$

\Ex{simpmod} Let us consider the ``triangular'' or \is{simplicial}
polynomial subspaces
$$
\CT_{n} = \set{x^{i}y^{j}}{0 \leq i+j \leq n }.  \Eq{Tn}
$$
We remark that $\CT_n$ forms a module for the standard
representation of the Lie algebra $\sL3$ by first-order differential
operators, \crf{GKO}, and so plays an important role in the theory of quasi-exactly solvability
and orthogonal polynomials, \crf{Turbpoly,Turbpolz}. 
It is not hard to see that $\CT_n$ forms a regular subspace of order $s=n$.
\pari
To construct the dual
basis, we look for a set of differential operators
$L_{ij}, 0\leq i+j \leq n$, such that 
$$
L_{ij}(x^{k}y^{l}) = \delta_{ik} \delta_{jl}, \forall
0 \leq i+j \leq n, \quad 0 \leq k+l \leq n. \Eq{Lijdef}$$
Let 
$$\dop = x\,\partial _x+y\,\partial _y$$
 denote the degree or scaling operator. Let
$$p_k(x) = {(-1)^k\over k!}\,(x-1)(x-2)\cdots (x-k)$$
be the unique  polynomial of degree $k$ such that
$p_k(0)=1$ and $p(j)=0$ for $\rg jk$.  Then dual basis operators are given by
$$L_{ij}= \fra{i!\,j!} \> p_{n-i-j}(\dop) \cdot \partial _x^i\,\partial _y^j.  \Eq{Lij}$$
Indeed, the two last factors annihilate any monomial of degree $\leq i+j$ except for
$x^iy^j$, whereas the polynomial in $\dop$ will annihilate the higher degree
monomials.  For $n=2$, the dual basis is explicitly given by
$$\eeq{
L_{20}=\f2 \partial _x^{2},\qquad 
L_{11}=\partial _{xy},\qquad 
L_{02}=\f2 \partial _y^{2}, 
\\
L_{10}=(-\dop+1)\cdot \partial _x = (-x{\partial_{x}}-y{\partial_{y}}+1)\cdot \partial_{x},\\ 
L_{01}=(-\dop+1)\cdot \partial _y = (-x{\partial_{x}}-y{\partial_{y}}+1)\cdot \partial_{y},\\
L_{00}=  \f2\dop^2-\fr32\dop+1 =  \f2 x^{2}\partial_{x}^{2}
+xy{\partial_{xy}}+ \f2 y^{2}\partial_{y}^{2}-x{\partial_{x}}-y{\partial_{y}}+1.}
\Eq{T2dual}
$$
\eqF{Lij} readily generalizes to the simplicial
modules  in
$m$ variables, \ie the polynomial subspaces generated by the monomials
$$\qeq{x^I = x_{1}^{i_1}\cdots x_{k}^{i_k}, \\\#I = \sum_{l} i_{l} \leq n. } \Eq{Tnm}$$
We note that this subspace forms a finite-dimensional module for the standard representation of $\sL
n$ by first-order differential operators.

The proof of \th{dM} ultimately rests on the following technical
Lemma.

\Lm{A} Let $A(x)$ be a real-analytic, $r\times n$ matrix-valued
function, defined for $x\in X\subset \R^m$.  Suppose $\rank A(x) = r$
for all $x\in X$.  Let $b\colon X\to \R^r$ be an analytic
vector-valued function.  Then there exists a solution $h\colon X\to \R^n$ to the 
matrix equation $A(x) h(x) = b(x)$ which is
analytic for all $x$.

\Proof Clearly, it suffices to prove the result when $b(x) = e_j$ is
constant, equal to the \jth basis vector.  Let $h_j(x)$ denote the
sought-after solution with this particular choice of the right hand
side. Let $v_1(x),\ldots, v_r(x)$ denote the rows of $A$, considered
as vector-valued functions $v_k\colon X\to \R^n$. Let $\Hstar\colon
\Wedge ^k\Rn \to \Wedge ^{n-k}\Rn $ denote the flat space Hodge star
operation in the exterior algebra $\Wedge\Upstar\Rn$, \crf{Warner; p.
  79}.  In particular, $*1 = e_1\wedge \cdots \wedge e_n\in \Wedge
^n\Rn$ is the volume form. One can write the matrix system $A(x)
h_j(x) = e_j$ in the equivalent exterior form
$$v_i(x) \wedge [\Hstar h_j(x)] = \mcases{\,0\,,&i\ne
  j,\\\Hstar1\,,&i=j,}\Eq{vHstar}$$
obtained by applying $\Hstar$ to
each equation and using the fact that $v\cdot w = \Hstar( v\wedge
\Hstar w)$. An evident solution to the first $r-1$ equations in
\eq{vHstar} is
$$h_j(x) = \Hstar\left [\> v_1(x)\wedge \cdots \wedge v_{j-1}(x)\wedge
  v_{j+1}(x)\wedge \cdots \wedge v_r(x)\wedge\gamma (x) \>\right
]\,,\Eq{v1rm}$$
where $\gamma \colon X\to \Wedge ^{n-r}\R^n$ is an
arbitrary analytic map.  The final equation in \eq{vHstar} leads to
$$(-1)^{n+j}\,v_1(x)\wedge \cdots \wedge v_r(x)\wedge \gamma (x) = \Hstar 1.\Eq{final}$$
This is a linear equation in the coefficients $g_1(x),\ldots ,g_k(x)$ of
$\gamma (x)$, where $k = {n \choose r}$, and as such has the  form
$$a_1(x)g_1(x) + \cdots + a_k(x)g_k(x) \equiv 1, \Eq{corona}.$$ 
Here $a_1(x),\ldots ,a_k(x)$ are analytic functions explicitly determined
by $v_1(x),\ldots ,v_r(x)$; indeed, up to sign, they are just the rank
$r$ minors of $A(x)$.  Moreover, these minors cannot simultaneously
vanish since $v_1(x),\ldots ,v_r(x)$ are assumed to be everywhere
linearly independent. Since we are dealing with real-valued functions,
an evident solution to \eq{corona} is
$$g_\nu (x) = {a_\nu (x) \over a_1(x) ^2 + \cdots +a_k(x)^2}\>,\qquad
\rg \nu k. \Eq{realsol}$$
Substituting \eq{realsol} into \eq{v1rm} and using the fact that
$\Hstar(\omega \wedge \Hstar\omega) =\nnorm \omega ^2$ for any $\omega
\in \Wedge\Upstar\Rn$, produces an explicit analytic solution to
\eq{vHstar} in the form
$$h_{j}(x) = (-1)^{n+j}\Hstar\left [{ v_1(x)\wedge \cdots \wedge
    v_{j-1}(x)\wedge v_{j+1}(x)\wedge \cdots \wedge v_r(x) \wedge \mskip-3mu
    \Hstar\mskip-3mu\bigl(v_1(x)\wedge\cdots\wedge v_{r}(x)\bigr) \over
    \nnorm{v_1(x)\wedge \cdots \wedge v_{r}(x)}^2}\right ]\Eq{Aj}$$
Note that the denominator is nowhere zero since the rank of $A(x)$
equals $r$.  \qed

\lm{A} plays a critical role in our theory. It relies on the fact that there
exists an analytic solution $a_1(x),\ldots ,a_k(x)$ of \eq{corona}, which, in the
real category considered here, is constructed in \eq{realsol}. The advantage of choosing the
reals as the ground field lies in the fact that our results are valid for
arbitrary domains.  However, it is important to keep in mind that the solutions to
\eqe{corona} given by
\eq{realsol} may fail to be analytic in the complex category, and indeed, \lm A
is not true for all complex domains $X$. 

\Ex{AC} A simple counterexample is
provided by the case $X = \Cx2
\setminus \{(0,0)\}$, and the $1\times 2$ matrix function $A(x,y) = (x,y)$, which has rank $1$ for
all
$(x,y)\in X$.  There is no complex-analytic vector-valued function $h(x,y) =
(h_1(x,y),h_2(x,y))^T$ such that
$$A(x,y)\> h(x,y) = x\,h_1(x,y) + y \,h_2(x,y) \equiv 1,\Eq{xyA}$$
for all $(x,y)\ne (0,0)$.
This stems from the fact that $X$ is not a domain of holomorphy, and so any
complex-analytic function on
$X$ can be extended to a complex analytic function on all of $\Cx2$, \crf{Krantz;
\S 0.3.1}.  Extending $h_1(x,y)$ and $h_2(x,y)$ in this manner, by continuity \eq{xyA} would also
have to hold at
$x=y=0$, which is clearly impossible.  

The preceding example makes clear
that the complex case is more subtle, and requires additional
assumptions.  One could, for instance, restrict
oneself to the case where $\F$ is the algebra of complex-analytic germs (\ie
convergent power series).  In this case, the natural analogue of the
order of an $r$-dimensional module would be the smallest $n$ such that
the $r\times n$ matrix formed by the constant terms of the \nth order Wronskian $\Wn$ has rank
$r$.  For such an $n$, one of the rank $r$ minors of $\Wn$ would be a
unit, thereby making \lm A true.

The utilization of more general domains $X\subset\Cn$ would require a
cohomological assumption in order that \eq{corona} admit an analytic
solution.  One could, for instance, demand the vanishing of every
$H^1$ whose coefficients lie in a coherent sheaf, or more generally
that $X$ be a Stein manifold, \rf{GunningRossi}.  Alternatively, one
could impose a geometric restriction on the domain $X$.  For example,
Corollary 7.2.6 in Krantz, \rf{Krantz; p. 293}, states that if
$X\subset \Cm$ is a \is{pseudoconvex} subdomain, then \eq{corona} has
a complex analytic solution $g_1(x),\ldots ,g_k(x)$ provided the
$a_\nu $'s do not simultaneously vanish.  Therefore, \lm{A}, and its
consequences, would hold provided $X$ is pseudoconvex.  To keep
matters simple, though, we shall not return to the complex-analytic
situation.

\Label{Proof of \th{dM}}
Recall first that we are seeking dual basis operators of order $s$.
To construct an \ith dual basis operator, we use the isomorphism
\eq{sig} to rewrite \eqe{dualops} in the equivalent matrix form
$$\W s(x)\,\bo h_i(x) = e_i,\Eq{dbW}$$ 
where $\bo h_i(x) = \sigma_s(L_i) \in \Fq s$ is the vector of coefficients of the operator
$L_i$, and $e_i$ is
the standard \ith basis vector for $\R^r$.  Since $s$ is the order of
the subspace $\M$, we have $\rank \W s(x) = r$ for all $x$, and hence
\lm{A} shows that there exists a solution $\bo h_i(x)$ to \eq{dbW}
which is analytic for all $x\in X$.  This simple construction produces
the required basis for the affine annihilating operators.  \qed

Having proven the existence of a dual operator basis, we can return to
the examination of the annihilator ideal.  Thus, for the remainder of
this section we assume that $\M$ is a regular $r$-dimensional subspace
of functions with basis $f_1,\ldots, f_r$, and fix a dual basis
$L_1,\ldots, L_r\in \KM$ of operators of order $s$ or less.  It is important
to keep in mind that $\dM$, as defined, is a vector space, and not an
$\F$-module.  When the need arises, we will use $\dMtilde$ to denote
the $\F$-module generated by $L_1,\ldots, L_r$.  Our results rely on the following key observation,
whose straightforward proof is left to the reader.

\Pr{annproj}
The mapping $\a\colon \D\to \A$ given by
$$\a(T)=T-\sum_{\kappa=1}^r T[f_\kappa]\, L_\kappa,\qquad \qquad  T\in \D, \Eq{annproj}$$
defines an $\F$-module homomorphism.  Furthermore,
$\a$ is a left inverse of the inclusion homomorphism $\iota\colon \A\to \D$, thereby yielding the
$\F$-module decomposition $\D \simeq \A\>\oplus\dMtilde$.

At this point it is also important to note that a basis $f_1,\ldots
f_r$ of $\M$ does not uniquely determine the dual basis $L_1,\ldots,
L_r$; one can obtain other bases by adding elements of $\A^s$ to the
operators $L_\kappa$.  For this reason the projection
$\a\colon \D\rightarrow\A$ is not a natural object, but rather depends on
the choice of the dual basis.

\Remark On the other hand, for a simplicial module $\CT_n$ described in \ex{simpmod}, there are no
annihilators of degree less than
$n+1 = s+1$, and hence the dual module $\dM$ is \is{uniquely} determined, or equivalently, $\A^s=0$.
Thus, the splitting of $\D \simeq \A\>\oplus\dMtilde$ described in \pr{annproj} is canonical in
this case. Subspaces having this property will be called \is{saturated}, and form an interesting
class worth further investigation.

For each multi-index $I$ let us denote the ``basic'' annihilator
$$K_I = \a(\partial_I) \Eq{KI}$$
obtained by projecting the basis differential operators $\partial _I$ onto the annihilator $\A$. 
It is  important to note that since the $L_\nu $ have orders at most $s$,
if $\#I \geq s+1$, then the leading order term (or symbol) of $K_I$ is just
$\partial_I$.

\Co{annlc} Any differential operator $T\in \D$ can be written as a finite linear combination
$$ T = \sum _{\nu =1}^r g_\nu (x)\, L_\nu  + \sum _I b_I(x) \,K_I, \Eq{annlc}$$
of the dual affine annihilators and the basic annihilators.

\Proof  Since the operators $\partial_I$ generate $\D$ as
an $\F$-module, we may use \pr{annproj} to infer that $L_1,\ldots,L_r$
along with the operators $K_I$ also generate $\D\,$; and that the affine
annihilators along with all $K_I$ such that $\# I\leq s$ generate
$\D^s$.  It follows immediately that for $n\geq s$, the $K_I$ such
that $\# I\leq n$ generate $\A^n$. \qed

\Remark The linear combination \eq{annlc} is not necessarily unique.
However, one can eliminate precisely $r$ of the $K_I$'s in order to
suppress any redundancy and thereby yield a well-defined $\F$-module
basis $\{L_\nu ,K_I\}$ for $\D$.

\Th{highann}  Let $\M\subset\F$ be a regular $r$-dimensional subspace of order
$s$.  Then its annihilator ideal $\AM$ is finitely generated by differential operators
of order at most $s+1$.

\Proof  We shall prove that the operators $K_I$ such that $\# I\leq s+1$
generate $\CA$ as an ideal of $\D$.  Clearly all of $\CA^{s+1}$ can be
so generated.  Let $J$ be a multi-index whose order is $\#J > s+1$, and
choose multi-indices $I, N$ with $\# I = s+1$ and such that
$\partial_J = \partial_{N}\partial_I$.  According to the remark at the end of
the proof of \co{annlc}, $K_J$ and $\partial_{N}\cdot K_I$ have the same
leading term, and therefore the order of the difference $K_J - \partial _N\cdot K_I$ of the two
operators will be smaller than the order of $J$.  The desired
conclusion now follows by induction. \qed

\Remark The set of generators $K_I$ constructed above is typically not
minimal.  However, since the set in question is finite, minimal
generating sets do exist.

\Remark  In the theory of $D$-modules, one is interested in subspaces generated by rational
functions  $f(x)=p(x)/q(x)$, with polynomial $p,q$. However, the annihilating operators of interest
are required to have polynomial coefficients, and so the set-up is a bit different from that
considered in this paper.  
 See \rf{OakuTak} for applications of powerful techniques from Gr\"obner bases and the theory
of $D$-modules towards the determination of the annihilators of such rational subspaces.
It would be interesting to see whether our techniques have anything to add to this theory.
 
\Ex{cosx}  Consider the two-dimensional subspace $\M = \{1,\cos x\}$.  The second order Wronskian
matrix is
$$\W2(x) = \pmatrix{1&0&0\cr\cos x&-\sin x&-\cos x\cr},$$
and hence
$\M$ is regular of order $2$.  Applying the algorithm of \lm{A} leads to the
dual operators
$$\qeq{L_1 = \cos^2 x \,\partial _x^2 + \cos x \sin x \,\partial _x + 1,\\
  L_2 = -\cos x \,\partial _x^2 - \sin x \,\partial _x.}$$
It's
important to mention that often the algorithm does not produce the
most efficient answer.  Indeed, for this particular $\M$, a more
stream-lined basis of dual operators is given by
$$\qeq{
  L_1 = \partial_x^2+1,\\
  L_2 = -\cos x \,\partial _x^2 - \sin x \,\partial _x.}
$$

The annihilator ideal, $\A(\M)$ is generated as a ring by the
following operators:
$$A_1=-\sin x\,\partial_x^2+\cos x\,\partial_x,\quad
A_2=\partial_x^3+\partial_x.$$
This is a consequence of \th{highann}, which states that $\A(\M)$ is
generated by the projections of the order $3$ basic differential
operators via \eq{annproj}.  Indeed the projected operators are:
$$
\a(1)  = \sin x\, A_1,\quad
\a(\partial _x)  = \cos x\, A_1,\quad
\a(\partial _x^2)  = -\sin x\, A_1,\quad
\a(\partial _x^3)  = A_2-\cos x\, A_1
$$
Therefore $A_1$ and $A_2$ suffice to generate all of $\A(\M)$.

\Ex{xcos2x} The subspace $\M = \{1,\cos x,\cos 2x\}$ considered in
\ex{cos2x} is considerably more difficult, since one needs to use the
fourth order Wronskian to implement the algorithm.  The explicit
formulae are quite complicated.  Set
$$\nu (x) = {144 - 432 \sin ^2 x + 432 \sin ^4 x + 400 \sin ^6 x}.$$
Indeed, $\nu(x)$ is the sum of the squares of the rank 3 minors of the
fourth order Wronskian.  The reciprocal of $\nu(x)$ will therefore be
a factor in the expressions for the dual operator basis obtained from
the algorithm of \lm{A}.

The dual operators are given by
$$\eeq{ \nu (x) L_1 = \nu (x) + (180 \cos x \sin x - 612 \cos x \sin
  ^3 x + 696 \cos x \sin ^5 x) \partial _{x} \creq + (180 - 720 \sin
  ^2 x + 1164 \sin ^4 x - 616 \sin ^6 x) \partial _x^2 \creq + (36
  \cos x \sin x + 180 \cos x \sin ^3 x - 456 \cos x \sin ^5 x)
  \partial _x^3 \creq + (36 - 144 \sin ^2 x - 60 \sin ^4 x + 136
  \sin ^6 x)
  \partial _x^4 , \\
  \nu (x) L_2 = (-192 \sin x + 656 \sin ^3 x - 736 \sin ^5 x) \partial
  _{x} \creq + (-192 \cos x + 576 \cos x \sin ^2 x - 656 \cos x \sin
  ^4 x) \partial _x^2 \creq + (-48 \sin x - 176 \sin ^3 x + 496 \sin
  ^5 x) \partial _x^3 \creq + (-48 \cos
  x + 144 \cos x \sin ^2 x + 176 \cos x \sin ^4 x) \partial _x^4 , \\
  \nu (x) L_3 = (12 \cos x \sin x - 20 \cos x \sin ^3 x) \partial _{x}
  + (12 - 24 \sin ^2 x + 20 \sin ^4 x) \partial _x^2 \creq + (12 \cos
  x \sin x + 20 \cos x \sin ^3 x) \partial _x^3 + (12 - 24 \sin ^2 x
  - 20 \sin ^4 x) \partial _x^4 , }$$
Once again, since we are not
dealing with a saturated module, the choice of the dual operators is
not canonical, and there exists a more stream-lined dual basis:
$$\eeq{
L_1=\f{4}\,\partial_x^4+\fr5{4}\,\partial_x^2+\partial_x, \\
L_2=-\f{3}\cos x\,\partial_x^4 - \f{3}\sin x\,\partial_x^3
-\fr4{3}\cos x\,\partial_x^2-\fr4{3}\sin x\,\partial_x\\
L_3=\f{12}\partial_x^4+\f{6}\sin x\,\cos x\,\partial_x^3
+\left(\f{12}+\f{2}\sin^2 x\right)\partial_x^2-\f{3}\sin x\,\cos x\,\partial_x
}$$

The annihilator ideal, $\A(\M)$ is generated as a ring by the
following operators:
$$\eeq{ A_1=4(\cos^2x -1)\,\partial_x^3 + 12\sin x\cos x\,
  \partial_x^2-4(2\cos^2x+1)\,\partial_x \\
  A_2= -\sin x\,\partial_x^4+\cos x\,\partial_x^3-4\sin
  x\,\partial_x^2+4\cos x\,\partial_x \\
  A_3=\partial_x^5+5\,\partial_x^3+4\,\partial_x }$$
To confirm this one
needs to check that $A_1, A_2, A_3$ generate the projections
$\a(\partial_x^n),$ $n=0,\ldots,5$.  Indeed the projected operators
are given by:
$$\eeq{
\a(1) = -\f{24}\sin 2x\,A_1+\f{6}\sin x\,A_2,\quad
&\a(\partial_x) = -\f{12}\cos 2x\,A_1,\\
\a(\partial_x^2) = \f{6}\sin 2x\,A_1+\f{3}\sin x\,A_2,\quad
&\a(\partial_x^3)= \f{3}\cos 2x\,A_1+\cos x\,A_2,\\
\a(\partial_x^4) = -\fr2{3}\sin 2x\,A_1-\fr7{3}\sin x\,A_2,\quad
&\a(\partial_x^5)= -\fr4{3}\cos 2x\,A_1-5\cos x\A_2+A_3
}$$

\bigskip

 \Section{non} Nonlinear Operators and Invariant Subspaces.

Let us next turn to the characterization of nonlinear annihilating and
affine annihilating operators.  We let $\E = \EX$ denote the space of
nonlinear differential operators.  More specifically, in the case of
analytic functions, $\F = \CaX$, the space $\E$ consists of all
analytic differential functions, \crf E, meaning analytic functions
$F\colon \Jn(X,\R) \to \R$ defined (globally) on the \nth order jet
space of real-valued functions on $X$. Here $n\geq 0$ is the order of
the differential function $F$.  We write $F[u] = F\xun$ for such an
operator, where the square brackets indicate that $F$ depends on $x,u$
and derivatives of $u$.

Our main goal is to prove a structure theorem for those differential operators, both linear and
nonlinear, which preserve a given subspace $\M\subset \F$.  Throughout
this section we assume that $\M$ is a regular $r$-dimensional subspace of order $s$, with
basis $f_1,\ldots, f_r$ and dual basis $L_1,\ldots, L_r$. 
Let $\NM \subset \E$ denote the set of nonlinear annihilating
operators, \ie those operators that map every function in $\M$ to
zero.  Using our basic annihilating operators \eq{KI}, we can readily construct the most general
nonlinear annihilating differential operator for our subspace.

\Th{NA} Every operator $F\in \NM$ can be written as a finite sum 
$$F[u] = \sum_{I} \ G_I[u] \cdot K_I[u], \Eq{NA}$$
where the $G_I$ are arbitrary elements of $\>\E$ and $K_I = \a(\partial_I)$ are the basic
annihilating operators.

\Proof From \co{annlc}, we know that there is an analytic
function $G(x,y,z)$, where $y = \psubs yr$, $z = (\,\ldots ,z_I,\ldots\,)$, such that
$$F[u] = G(x,L_1[u],\ldots ,L_r[u], \ldots , K_I[u],
\ldots\;), \Eq{FG}$$
depending on finitely many of the $K_I$.  If $u = \sum c_i\,f_i(x)
\in \M$, then $L_i[u] = c_i \in \R$, and so substituting $u$ into \eq{FG} yields
$$F[u] = G(x,c,0) = 0, \where c = \psubs cr.\Eq{G0}$$
\eqE{G0} will hold for all $x,c$ if and only if 
$$G(x,y,z) = \sum_I G_I(x,y,z) \,z_I.$$
Therefore,
$$G(x,L_1[u],\ldots ,L_r[u], \ldots , K_I[u], \ldots\;) = \sum_{\nu =
1}^q G_I[u] \,K_I[u],$$
where the coefficients $G_I[u]$ are differential functions.  \qed

Let $\subs Kl$ be a (minimal if desired) generating set for the annihilator $\A$.
For example, according to \th{highann}, one can choose the $K_\nu $ from among the basic
annihilators $K_I$ of orders $\#I \leq s+1$.  Then we can immediately simplify \eq{NA} to only use
the generating annihilating operators.

To be precise, let $\G = \GX$ denote the $\E$-module consisting of
differential operators whose coefficients are differential functions.
Such an operator $Z \in \G$ is given by a finite sum
$$Z = \sum _I \ G_I[u]\> \partial _I = \sum _I\  G_I\xun \>\partial _I\>.
\Eq{Gdop}$$ 
Note that the operator $G\colon \E\to \E$ maps differential functions to differential
functions.

\Co{NAmin} Every nonlinear annihilating operator $F\in \NM$ has the form 
$$F = \sum_{\nu =1}^l \ Z_\nu \cdot K_\nu , \Eq{NAmin}$$
where $\subs Zl$ are arbitrary elements of $\>\G$.

We are now in a position to prove the main result of the paper.  
Let
$$\qeq{\PM = \set{P\in \D}{P[\M]\subset \M},\\
\QM = \set{Q\in \E}{Q[\M]\subset \M}}
\Eq{PQM}$$
denote, respectively, the left ideals consisting of all linear, respectively nonlinear,
differential operators that preserves the given subspace $\M$.
 
\Th{PQM} Let $\M\subset \F$ be a regular $r$-dimensional subspace of analytic functions of order
$s$.  Let $\subs Lr\in \KM$ be a dual basis for its affine annihilator, and
$\subs Kl\in \AM$ a generating set of annihilating operators.  Then
every nonlinear operator $Q\in \QM$ that preserves $\M$ can be
written in the form
$$Q[u] = \sum _{i=1}^r \; f_i(x)\, H_i(L_1[u],\ldots ,L_r[u]) + \sum_{\nu =1}^l\; Z_\nu[u] \cdot K_\nu
[u], \Eq Q$$ 
 where the $H_i\in \Ca(\Rx r)$ are arbitrary analytic functions, and where the
$Z_\nu \in \GX$ are arbitrary operators.

\Proof
Suppose $f(x) = \sum c_j f_j(x)\in \M$.  Since $Q[f] \in \M$, for each $\rg ir$ we have
 $L_i[Q[f]] = H_i(c_1,\ldots ,c_r)$ is a constant depending on the coefficients of $f$. 
It follows immediately that 
$$Q[u] - \sum _{i,j=1}^r \sum _{i=1}^r \; f_i(x)\, H_i(L_1[u],\ldots ,L_r[u]) = 0$$
for all $u\in\M$, and so the result \eq Q follows immediately from
\eq{NAmin}.  \qed

In particular, every linear operator $P\in \PM$ that preserves $\M$ can be
written in the form
$$P = \sum _{i,j=1}^r a_{ij} f_i L_j + \sum_{\nu =1}^l R_\nu \cdot K_\nu ,\Eq P$$ 
where the $a_{ij}\in \R$ are arbitrary constants, and the
$R_\nu \in \D$ are arbitrary linear differential operators.
Therefore, we conclude:

\Co{Pdecomp} The space of linear differential operators leaving a regular subspace $\M\subset \F$ of
analytic functions invariant can be decomposed into a semi-direct product $\PM \simeq \EndM
\semidirect \AM$ of the space $\EndM = \{A\colon \M\to \M\} \simeq \M \tensor \dM$ of linear
endomorphisms with the annihilator ideal $\AM$.

In the case when $\M$ is generated by polynomials, this corollary is proved directly by Post and
Turbiner, \rf{PostTurb}.  Thus, the class of regular subspaces form the proper analytic
generalization of the polynomial subspaces.  See also \rf{FKp, PostHijl}.

In the one-dimensional case, the expressions $L_i[f]$ are directly related to the
generalized symmetries considered by Svirshchevskii, \rf{Sv}.  The
following example serves to illustrate this.

\Ex3  Let $p=1$ and let $\M$ denote the space of quadratic polynomials $ax^2 + b x + c$ in the
scalar variable $x$, with basis $1,x,x^2$. The annihilator is generated by $K = \partial _x^3$
since the subspace forms the solution space to the linear ordinary differential equation
$$u_{xxx} = 0. \Eq{uxxx0}$$
 According to
\rf{Sv; Example 3}, the fundamental invariants for \eq{uxxx0} are $I_\nu(x,u\ps2)  = J_\nu [u]$,
where
$$\qeq{J_1 = \partial _x^2, \\J_2 = x\partial _x^2 - \partial _x, \\
J_3 = x^2\partial _x^2 - 2x\partial _x + 2.}$$
Since
$$\eeq{J_1[1] = 0,&J_2[1] = 0,&J_3[1] = 2,\\ 
J_1[x] = 0,&J_2[x] = -1,&J_3[x] = 0,\\
J_1[x^2] = 2,&J_2[x^2] = 0,&J_3[x^2] = 0,}$$
we see that $L_1 = \f2J_3, L_2 = -J_2,L_3 = \f2J_1$ are the dual operators with respect to the
given basis of $\M$.  In accordance with the general result, every nonlinear differential operator
$Q\in
\QM$ leaving
$\M$ invariant takes the form
$$Q[u] = A_0[u] + A_1[u] x + A_2[u] x^2 + T[u] \cdot K[u],$$
where 
$$\eeq{A_\nu[u] = H_\nu (J_1[u],J_2[u],J_3[u]) = H_\nu ( u_{xx}
, x u_{xx} - u_x, x^2 u_{xx} - 2 x u_x + 2 u),}$$
with $H_\nu\in\Ca(\Rx 4)$ arbitrary, and where $T\in \G$ is an arbitrary (nonlinear)
differential operator.

The precise connection between our approach and that of Svirshchevskii relies on the
 theory of generalized symmetries of differential equations, as presented, for example, in
\rf{O; Chapter 5}.  We assume that the reader is familiar with this theory for the remainder of this
section, and take $u$ to be a scalar variable, although vector-valued generalizations are
straightforward.

\Pr{lgs}  Let $\Delta [u] = 0$ be an analytic, homogeneous system of linear differential equations
and let
$\M\subset \F$ denote the vector space of solutions. Then the generalized vector field $\v_Q =
Q[u]\,\partial _u$ is a symmetry of
$\Delta =0$ if and only if the differential operator $Q[u]\in \QM$ leaves $\M$ invariant.

Note that $Q[u]$ may be nonlinear, although many nondegenerate linear partial differential equations
only admit linear generalized symmetries; see \rf{ShaShi}.  \pr{lgs} is an immediate consequence of
the basic definition of generalized symmetry, which requires that the vector field $\v_Q$ leave the
solution space to $\Delta =0$ infinitesimally invariant.  Linearity of the differential equation
implies that infinitesimal invariance coincides with invariance of the solution space.  Thus, if the
differential equation is of finite type,
\rf{GJ}, its solution space is finite-dimensional, and hence we can completely characterize the
generalized symmetries of the system using the affine annihilators of the solution space along with
\th{PQM}.

In Svirshchevskii's method, one considers an $r$-dimensional  subspace $\M$ consisting of analytic
functions of a single variable $x$. Assume, for simplicity, that $\M$ is regular of order $r-1$. 
The linear ordinary differential equation that characterizes $\M$ is given by
$$K_r[u] = 0,\Eq{Kr0}$$
where $K_r$ is the \rth order differential operator \eq{Kr} constructed in \ex{AR}.  The resulting
formulae \eq Q for the generalized symmetries $\v_Q$ of \eq{Kr0} is written in \rf{Sv} in terms of
the first integrals of \eq{Kr0}.
Recall that a differential function  $F(x,u\ps{r-1})$ depending on at most \xsts{r-1}order
derivatives of $u$ is called a
\is{first integral} of the \rth order ordinary differential equation \eq{Kr0} if and only if its
derivative $D_xF = 0$ vanishes on the solutions.  The first integrals of a linear ordinary
differential equation can be constructed in terms of the solutions to the adjoint equation
$$K_r\Upstar [z] = 0,\Eq{Kr9}$$
where $K_r\Upstar$ is the usual (formal) adjoint differential operator, \crf{O; p. 328}.  Indeed,
integration by parts shows that if $z(x)$ is any solution to the adjoint \eqe{Kr9}, then
$$z\, K_r[u] = D_x F[z,u] \Eq{1i}$$
determines a first integral to \eq{Kr0}.  The explicit formula for the first integral $F[z,u]$ is
classical, and described in \rf{Sv}.

On the other hand, since a first integral is constant on the solution space $\M$ to \eq{Kr0}, it
\is{is} an affine annihilator.  Choosing a basis $\subs zr$ for the solution space to the adjoint
equation
\eq{Kr9} produces $r$ linearly independent first integrals, and hence a basis for the operator
dual of the subspace.  Thus, in this manner we recover  Svirshchevskii's construction of the
generalized symmetries of homogeneous, linear ordinary differential equations.

Conversely, given a nonlinear ordinary differential equation, $Q[u] = 0$, one can construct the
subspaces $\M\subset \F$ it leaves invariant by the following ``inverse symmetry procedure''. 
One needs to classify all linear ordinary differential equations which admit
$\v_Q$ as a generalized symmetry.  For example, it can be shown that if $Q(x,u\ps n)$ is a
\iz{nonlinear} generalized symmetry of a nonsingular
linear ordinary differential equation \eq{Kr9} of order $r$, then, necessarily $r \leq 2n + 1$. 
Consequently,  any invariant subspace of a nonlinear \nth order differential
operator has dimension at most $2n+1$. In practice, the determination of the equations of
a prescribed order possessing a given generalized symmetry is a straightforward adaptation of the
usual infinitesimal computational algorithm for symmetry groups of differential equations, \crf{O}. 
Examples of this procedure appear in \rf{Sv,Svx,Svi}. 

For functions depending on more than one independent variable, a similar procedure works, although
now one can no longer use a single linear partial differential equation to characterize the
subspace, but must employ a linear basis for the annihilator, which will form a linear systems of
partial differential equations of finite type. The outline of this method is reasonably clear, but
full details remain to be worked out. In particular, the existence of corresponding bounds on the
dimensions of invariant subspaces for nonlinear partial differential operators is not known.

\Section{app}  Applications to Differential Equations.

A principal application of our theory is to find explicit solutions to both linear and nonlinear
differential equations.  In the theory of quasi-exact solvability, one considers an eigenvalue
problem
$$Q[u] = \lambda u, \Eq{Se}$$
which, in physical applications, is the stationary Schr\"odinger equation.  A linear differential
operator $Q$ is said to be \is{quasi-exactly solvable} if it leaves a finite-dimensional subspace
of wave functions $\M\subset \F$ invariant, so that $Q[u] \in \PM$.  
In this case, restricting to the subspace $\M = \{f_1(x),\ldots ,f_r(x)\}$, whereby
$$u(x) = c_1 f_1(x) + \cdots c_r f_r(x),\Eq{c}$$
reduces the eigenvalue problem \eq{Se} to a linear eigenvalue problem
$$A \, c = \lambda  \, c, \roq{for} c = \psubs cr^T.\Eq{Ac}$$
In this way, one reduces the solution of the differential equation \eq{Se} to an algebraic problem.
More generally, one can consider nonlinear differential operators $Q[u] \in \QM$ that leave $\M$
invariant, in which case the restriction of \eq{Se} to the subspace leads to a system of nonlinear
algebraic equations
$$H(c) = \lambda c \Eq{Hc}$$
for the coefficients in \eq{c}.

For evolution equations, 
the basic idea underlying the method of nonlinear separation goes back
to Galaktionov and his collaborators, \rfr{Galak-GPS2}, King, \rf{King}, and Fushchych
and Zhdanov, \rf{Fush95,FushZh}.  The basic idea appears in the following theorem.

\Th{eeQ}  Consider an evolution equation  
$$u_t = Q[u]. \Eq{ee}$$
Suppose the right hand side  $Q[u] \in \QM$
preserves a finite-dimensional  subspace $\M$, that is 
$$Q[u] = \sum _{i=1}^r H_i(L_1[u],\ldots ,L_r[u])\, f_i + \sum_{\nu =1}^l Z_\nu[u] \cdot K_\nu [u],
$$ 
 where the $H_i\in \Ca(\Rx r)$ are arbitrary analytic functions, and where the
$Z_\nu \in \GX$ are arbitrary operators. Then there exist ``separable
solutions'' of the evolution equation taking the form
$$u(x,t) = \sum _{i=1}^n \varphi _i(t) f_i(x), \Eq{sep}$$
if an only if  the coefficients $\varphi _1,\ldots ,\varphi _n$ are solutions to the dynamical system 
$$\od {\varphi _i}t = H_i\psubs \varphi n,\qquad \rg in. \Eq{dys}$$

The proof is immediate from \th{PQM}.  More generally, one can replace the evolution equation
\eq{ee} by a time-dependent Schr\"odinger equation 
$$i\, u_t = Q[u],$$
 resulting a complex first
order dynamical system, or a wave-type equation 
$$u_{tt} = Q[u],$$
which reduces to a second order dynamical system, with $\odw {\varphi _i}t$ on the left hand side
of \eq{dys}.  Indeed, one can apply the method to dynamical  partial differential equations
 of the general form 
$$T[u]=Q[u],\Eq{dpde}$$
 where $T$ is a linear ordinary differential operator in $t$, or, even more generally, a
time-dependent linear combination of operators $T_k(\pds utk)$ where each $T_k\in \QM$ leaves $\M$
invariant. An interesting class of examples are the equations of \is{Fuchsian type}, studied in depth by
Kichenassamy, \rf{KFuchs} in connection with blow-up and Painlev\'e expansions, \rf{KSr}, in which
$T[u]$ is a constant coefficient polynomial in the scaling operator
$S = t \partial _t$. 

\Remark See Cherniha, \rf{Cherniha}, for an even more general nonlinear separation ansatz in which
the basis functions
$f_i(x,t)$ (and hence the module) are also allowed to depend on $t$. 

\Ex{Tnee} We can already give many examples of non-linear QES evolution
equations in two space variables by using the generators \eq{Lij}.
Consider for example the simplicial subspace 
$${\cal T}_{2} = \{1,x,y,x^2,xy,y^2\}.$$ The dual basis of operators is 
given in \eq{T2dual}. Therefore, the most general second order ${\cal T}_{2}$ invariant evolution
equation  takes the form
$$u_t = A + B x + C y + D x^2 + E x y + F y^2,$$
where $A,B,C,D,E,F$ are arbitrary (linear or nonlinear) functions of the dual operators
$$\ceq{u_{xx},\quad u_{xy},\quad u_{yy},\quad u_x - x u_{xx} - y
  u_{xy},\\ u_y - x u_{xy} - y u_{yy},\quad \f2 x^2 u_{xx} + xy u_{xy}
  + \f2 y^2 u_{yy} - x u_x - y u_y + u.}$$
Higher order evolution
equations are obtained by adding in arbitrary annihilators.  Replacing
$u_t$ by $u_{tt}$ or other types of linear temporal differential
operators, leads to wave and more general types of equations that
leave the indicated subspace invariant. It follows, for example that
the evolution equation
$$
u_{t} = \f4\,x^{2}u_{xx}^{2}-xyu_{xy}-y^{2}u_{yy}+ \f8\,y^{2}u_{yy}^{3}
$$
admits the solutions given by
$$
u(x,y,t)
={x^{2}\over{k_{1}-t}}+k_{2}e^{-t}\,xy+{\sqrt{2}\over \sqrt{1-k_3e^{4t}}}\,y^{2}+k_3+k_4x+k_5y,
$$
where the $k_{i}$ are arbitrary constants determined by the initial
conditions. This is a very simple example which was chosen such
that the dynamical system governing the $c_{i}(t)$ was decoupled 
and therefore integrable by quadratures.

\def\va{\vec{\alpha}}
\def\vx{\vec{x}}
\Ex{99}
As a more substantial example, consider the following rotationally invariant evolution equation 
$$
\qeq{u_t = (\Delta-B)[\nabla(u^\sigma\,\nabla u) -Au^{\sigma+1}]+Cu+d,\quad
u=u(\vx,t),\\\sigma=1,2}\Eq{Req}$$ 
in $n$ space variables $\vx = \psups xn$.
We claim that the subspace 
$$\M=\{1,u_1=e^{\va_1\cdot\vx},u_2=e^{\va_2\cdot \vx}\}$$
 is invariant for a suitable choice of $\va_1,
\va_2$. To see this let us define
$$R[u] = \grad(u^\sigma\grad u)-A\,u^{\sigma+1},$$
and note the following identities:
$$
R[u+v]=R[u]+R[v]+\mcases{(\Delta-A)[uv],& \sigma=1,\\
(\Delta-A)[uv(u+v)],& \sigma=2.}
$$
Also let us note that for $\va\in \R^n$, 
$$R[e^{\va\cdot\vx}] = \bigl[(\sigma+1)\|\va\|^2-A\bigr]\> e^{\va\cdot\vx}.$$
Hence, taking $\va_1, \va_2$ so that $\|\va_{i}\|^2=A/(\sigma+1), \, i=1,2$, we will have
$$R[c_1 u_1]=R[c_2 u_2] = 0, \quad \forall c_1,c_2\in \R.$$
Of course $R$ does not annihilate all of $\M$; we are left with quadratic cross-terms.
Indeed, taking $u=c_1 u_1+c_2 u_2$ we have for $\sigma=1$
$$R[u] = R[c_1 u_1]+R[c_2 u_2] + c_1 c_2 (\Delta-A)[u_1 u_2].$$
Hence, taking $B=\|\va_1+\va_2\|^2$ will ensure that
$$(\Delta-B)[R[u]] = 0.$$
For $\sigma=2$, we need to take $B=\|2\va_1+\va_2\|^2$, which is the same as
$\|\va_1+2\va_2\|^2$.  Indeed, we have:
$$R[u] = c_1 c_2 (\Delta-A)[u_1 u_2 (u_1+u_2)] = C_1
(e^{(2\va_1+\va_2)\cdot \vx} + e^{(\va_1+2\va_2)\cdot x}),$$
and the right hand side is manifestly annihilated by $\Delta-B$.
It isn't hard to check that $R[u+c_3]$ differs from $R[u]$ by a linear combination of $u_1$,
$u_2$, and a constant.  Hence, $(\Delta-B)[R[u+c_3]]$ continues to lie in $\M$, and therefore
the right hand side of \eqe{Req} preserves $\M$.
\pari
We could also  consider the extended module 
$$\M = \{u_0=1, u_1=e^{\va_1\cdot \vx} , u_2=e^{\va_2\cdot \vx} ,
u_{-1}=e^{-\va_1\cdot \vx} , u_{-2}=e^{-\va_2\cdot\vx}\}.$$ 
The same reasoning as above can be applied to show that $R[u]$, where $u$ is a general
element of $\M$, is the sum of a linear combination of $R[u_i]$, $i\in\{1,2,-1,-2\}$ and a
number of cross-terms of the form constant times $u_i u_j$, $i\neq j$, $i,j\in \{0,1,2,-1,-2\}$.
We therefore require that $\Delta-B$ annihilate $u_1 u_2$, $u_1 u_{-2}$, and their reciprocals.
For $\sigma=1$, this constraint requires
$$B = 2 \|\va_1\|^2 = 2 \|\va_1\|^2, \andq  \va_1\cdot \va_2 = 0$$
so that $\va_1$ and $\va_2$ are perpendicular.
For $\sigma=2$, the conditions are:
$$B = 5 \|\va_1\|^2 = 5 \|\va_1\|^2, \andq  \va_1\cdot \va_2 = 0$$
and the conclusions are identical.
We remark that, by making use of the rotational invariance of $Q$, we can without loss of generality
assume that
$\va_1\cdot \vx=x^1$ and that $\va_2\cdot\vx =x^2$.

\Section z Conclusions.

In this paper, we have developed a general theory of invariant subspaces, and shown how these
results can apply to provide nonlinear separation of variables  ans\"atze for a wide variety of linear
and nonlinear partial differential equations.  These operators further provide a nonlinear
generalization of the quasi-exactly solvable operators of importance in the algebraic approach to
quantum mechanical systems.  A number of interesting problems warrant further development of this
method. 

\items{A significant mystery is the connection of this method with the
  Lie algebraic approach to quasi-exactly solvable modules.  For
  example, the linear operators preserving the simplicial subspace
  \eq{Tn} lie in the universal enveloping algebra of a standard
  realization of the Lie algebra $\glnR$, \crf{GKO, Turbis}.  However,
  this is not evident from the form of the affine annihilator and
  annihilator.  Thus, a fundamental issue is which subspaces admit
  such a Lie algebraic
  interpretation of the space of differential operators that leave them invariant.\\
  The ``inverse problem'' of characterizing the invariant
  finite-dimensional subspaces for a given linear or nonlinear
  operator is of critical importance.  The extension of
  Svirshchevskii's symmetry
  approach, \rf{Sv}, is under investigation.\\
  Applications of our analytical Wronskian methods to the algebraic
  context of $D$--module theory, \rf{BjorkD,Oaku,OakuTak}, looks quite
  promising.  In particular, the characterization of the analytic or
  polynomial annihilators of subspaces of rational functions would be
  a particularly
  interesting case.\\
  The formulae for the affine annihilators and annihilators are often
  extremely complicated, even for relatively simple subspaces.  (As an
  example, the reader is invited to write down the formulae for $\M =
  \{x^2,xy,y^2\}$.)  Moreover, in the final formulae \eqs PQ for the
  linear and nonlinear operators leaving the subspace invariant, one
  may encounter a significant amount of simplification and lowering of
  order.  Thus, it would be important to characterize low order
  operators in $\PM$ and $\QM$, as well as ``simple'' operators of
  physically relevant type, \eg elliptic, constant
  coefficient, Lorentz invariant, etc.\\
  Extensions of these methods to finite difference operators, building
  on the work of Turbiner, \rf{Turbis}, can be profitably pursued.}

\vskip 30pt

\Ack  We would like to thank the School of Mathematics of the University of
Minnesota for providing the travel funds that served to initiate this project.

\vskip 40pt

\References

\ends


%% file: o.tex


\hyphenation{di-men-s-ion-al quad-ra-tic 
   par-a-m-e-t-ri-ze par-a-m-e-t-ri-zes par-a-m-e-tri-zed 
   par-a-m-e-t-ri-z-a-tion par-a-m-e-t-ri-z-a-tions
   re-par-a-m-e-t-ri-ze re-par-a-m-e-t-ri-zes re-par-a-m-e-tri-zed 
   re-par-a-m-e-t-ri-z-a-tion re-par-a-m-e-t-ri-z-a-tions
   La-g-ran-gian La-g-ran-gians}


\magnification=\magstep1

\parskip = 1pt plus 2pt
\abovedisplayskip = 9pt plus 3pt minus 3pt
\belowdisplayskip = 7pt plus 3pt minus 3pt

\newskip\refparskip \refparskip = 1pt plus 1pt

\def\ignore#1{} \def\asis#1{#1}


\output{\ooutput}
\def\ooutput{\shipout\vbox{\makeheadline\pagebody\makefootline}%
  \advancepageno\global\fnotenumber=0
  \ifnum\outputpenalty>-20000 \else\dosupereject\fi}

\def\page{\vfill\eject}


\newif\ifwarn \warntrue

\def\msg#1{\immediate\write0{#1}}
\def\warn#1{\ifwarn\msg{** #1 **}\fi}
\def\warna#1{\ifwarn\msg{-> #1 <-}\fi}


\xdef\numcat{\catcode`0=11\catcode`1=11\catcode`2=11\catcode`3=11\catcode`4=11
\catcode`5=11\catcode`6=11\catcode`7=11\catcode`8=11\catcode`9=11\catcode`*=11}
\xdef\unnumcat{\catcode`0=12\catcode`1=12\catcode`2=12\catcode`3=12\catcode`4=12
\catcode`5=12\catcode`6=12\catcode`7=12\catcode`8=12\catcode`9=12\catcode`*=12}

{\numcat\gdef\0{0}\gdef\1{1}\gdef\2{2}\gdef\3{3}\gdef\4{4}\gdef\5{5}
\gdef\6{6}\gdef\7{7}\gdef\8{8}\gdef\9{9}
\gdef\10{10}\gdef\11{11}\gdef\12{12}\gdef\13{13}\gdef\14{14}\gdef\15{15}
\gdef\16{16}\gdef\17{17}\gdef\18{18}\gdef\19{19}
\gdef\20{20}\gdef\21{21}\gdef\22{22}\gdef\23{23}\gdef\24{24}\gdef\25{25}
\gdef\26{26}\gdef\27{27}\gdef\28{28}\gdef\29{29}
\gdef\30{30}\gdef\31{31}\gdef\32{32}\gdef\33{33}\gdef\34{34}\gdef\35{35}
\gdef\36{36}\gdef\37{37}\gdef\38{38}\gdef\39{39}
\gdef\40{40}\gdef\41{41}\gdef\42{42}\gdef\43{43}\gdef\44{44}\gdef\45{45}
\gdef\46{46}\gdef\47{47}\gdef\48{48}\gdef\49{49}
\gdef\91{A}\gdef\92{B}\gdef\93{C}\gdef\94{D}\gdef\95{E}
\gdef\96{F}\gdef\97{G}\gdef\98{H}\gdef\99{I}}

{\catcode`*=11\expandafter\gdef\csname *0\endcsname{0}
\expandafter\gdef\csname *1\endcsname{1}\expandafter\gdef\csname *2\endcsname{2}
\expandafter\gdef\csname *3\endcsname{3}\expandafter\gdef\csname *4\endcsname{4}
\expandafter\gdef\csname *5\endcsname{5}\expandafter\gdef\csname *6\endcsname{6}
\expandafter\gdef\csname *7\endcsname{7}\expandafter\gdef\csname *8\endcsname{8}
\expandafter\gdef\csname *9\endcsname{9}}

\def\hexnumber#1{\ifcase#1 0\or1\or2\or3\or4\or5\or6\or7\or8\or9\or
 A\or B\or C\or D\or E\or F\fi}
\def\romannumber#1{\romannumeral#1}
\def\alphanumber#1{\count255 = #1 \advance\count255 by 96 \expandafter\char\count255}

\newif\ifroman\romanfalse

\def\rnumbert#1{\ifroman{\tenit\romannumeral#1}\else{\tenrm\number#1}\fi}
\def\rnumberq#1{\ifroman"\romannumeral#1"\else\number#1\fi}


\def\Rough{\centerline{\bigit Rough\enskip Draft\enskip  \bigrm ---
\enskip \shortdate\enskip ---\enskip  
\bigit Do\enskip Not\enskip Distribute}\vglue 30pt}

\def\Title#1.{\def\\{\par}\ifproofmode\Rough\else\vglue 40pt\fi\atitle#1.\vskip20pt}

\def\Abstract#1\par{\par\bigskip\indent{\bf Abstract.}\enspace#1\par\bigskip}

\def\shortline#1{\hbox to 3truein{#1}}
\def\shorterline#1{\hbox to 2.5truein{#1}}

\def\boxits#1{\def\\{\hfil\egroup\par\shorterline\bgroup\asis}%
\vbox to 1.7in{\shorterline\bgroup#1\hfil\egroup\vfill}}
\def\boxit#1{\def\\{\hfil\egroup\par\shortline\bgroup\relax\ignorespaces}%
\vbox to 1.7in{\shortline\bgroup\ignorespaces#1\hfil\egroup\vfill}}

\newcount\autnumber  \autnumber=1

\def\email{\tt}
\def\www{\tt http://www.}
\def\author#1\\#2\support#3.{%
\expandafter\def\csname autx\romannumeral\the\autnumber\endcsname
   {\boxit{#1\fnotemark\\#2\\}
\vfootnote\fnotemark{\it Supported in part by #3.}
\global\advance\fnotenumber by 1}\global\advance\autnumber by 1}
\def\authorn#1\par{%
\expandafter\def\csname autx\romannumeral\the\autnumber\endcsname
   {\boxit{#1\\}}
\global\advance\autnumber by 1 }

\def\printauthor{\global\advance\autnumber by -1
\csname autform\romannumber\the\autnumber\endcsname}

\def\pjo{\author
Peter J. Olver\\
School of Mathematics\\
University of Minnesota\\
Minneapolis, MN\quad 55455\\
\email olver\@ima.umn.edu\\
\www math.umn.edu/$\sim$olver\\
\support NSF Grant DMS 98--03154.}

\def\pjoi{\author
Peter J. Olver\\
School of Mathematics\\
University of Minnesota\\
Minneapolis, MN\quad 55455\\
U.S.A.\\
\email olver\@ima.umn.edu\\
\www math.umn.edu/$\sim$olver\\
\support NSF Grant DMS 98--03154.}


\newcount\fnotenumber  \fnotenumber=0
\def\fnotemark{\ifcase\the\fnotenumber$^\dagmath$\or$^\ddagmath$\or
                         $^\Smath$\or$^\Pmath$\or$^\|$\or$\Upstar$\fi}
\def\fnote#1{{\footnote\fnotemark{\smallstyle #1\par\vskip-10pt}%
\global\advance\fnotenumber by 1}}

\skip\footins=20pt plus 4pt minus 4pt

\def\footnoterule{\hbox{\raise 5pt \vbox{\kern-3pt\hrule width 3 truein \kern 2.6pt}}}

\newdimen\qsize \qsize = 22 pc

\def\quote#1\qauthor#2\par{\medskip\noindent\hfil\vbox{\hsize=\qsize \smallstyle
#1\par\hfill --- #2\par\vfil}\hfill\par\medskip}


\def\amstexname{AmS-TeX}
\ifx\fmtname\amstexname
   \let\footnote\plainfootnote\def\cal{\fam2}%
   \def\>{\mskip\medmuskip}\def\=#1{{\accent22 #1}}%
    
\fi
\def\@{\char64 }

%
%

\newif\ifeqnwrite  \eqnwritefalse
\newif\ifeqnread  \eqnreadfalse
\newif\iflocal \localfalse

\def\keyout{\eqnwritetrue\immediate\openout1=\jobname.key }
\def\keyin{\openin2=\jobname.key \relax
\ifeof2 \warn{I can't find the file \jobname.key}\closein2 \else
\closein2 \numcat \input \jobname.key \unnumcat \fi}

\newtoks\outtoks
\def\wdef#1{\outtoks=\expandafter{\csname#1\endcsname}%
\immediate\write1{\global\def\the\outtoks{\csname#1\endcsname}}}


\outer\def\ends{\ifeqnwrite\closeout1\fi
\ifrefnum\closeout3\fi
\ifindex\closeout4\closeout5\closeout6\fi
\ifcontents\closeout7\fi
\vfill\supereject\end}

%
%

\newif\ifproofmode \proofmodefalse

\def\strutdepth{\dp\strutbox}
\def\margintagleft#1{\vadjust{\kern-\strutdepth\vtop to \strutdepth
   {\baselineskip\strutdepth\vss\llap{\sevenrm#1\quad}\null}}}
\def\prooflabel#1{\ifproofmode\margintagleft{#1}\fi}

\def\folio{\ifproofmode\jobname\quad\shortdate\quad\fi{\tenrm \number\pageno}}

\def\date{\ifcase\month\or January\or February\or March\or April\or May\or
June\or July\or August\or September\or October\or November\or December\fi
\space\number\day, \number\year}
\def\gobble#1#2{}
\def\shortdate{\number\month/\number\day/\expandafter\gobble\number\year}

\def\dated{\footnote\ {\vskip-3pt\hfill\bigrm \date}}


\def\firstpar{\relax}
\def\pari{\par}

\def\center{\parindent=0pt \parfillskip = 0pt 
\leftskip=0pt plus 1fil \rightskip = \leftskip \spaceskip=.3333em \xspaceskip = .5em
\pretolerance=9999 \tolerance=9999 \hyphenpenalty=9999 \exhyphenpenalty=9999\relax}

\def\atitle#1.{{\baselineskip=20pt \center \Bigrm #1\par\nobreak}}
\def\btitle#1.{{\center \bf #1\par\nobreak}}
\def\ctitle#1.{{\center \bigrm #1\par\nobreak}}

 \def\headfont{\bf} \def\Subheadfont{\it}

\def\Subhead#1{\medskip\heading{#1}\Subheadfont}

\def\headtagleft#1#2{\bigbreak{\parindent = 0pt\noindent\prooflabel{#2}\heading{#1}\headfont}}
\def\heading#1#2{{\noindent \rightskip = 0pt plus 1fil 
\spaceskip=.3333em \xspaceskip=.5em \hyphenpenalty=9999 
\exhyphenpenalty=9999 #2\hangindent=8pc #1\par\nobreak}\medskip}

%
%
%
%
%
%
%

\newcount\chapternumber  \chapternumber=1
\newcount\sectionnumber  \sectionnumber=1
\newcount\equationnumber  \equationnumber=1
\newcount\statementnumber \statementnumber=1
\newcount\figurenumber \figurenumber=1
\newcount\itemnumber \itemnumber=1

\newif\ifbeginningdocument \beginningdocumenttrue
\newif\iflongmode  \longmodefalse
\newif\ifbookmode \bookmodefalse
\newif\ifappendixmode \appendixmodefalse

\def\chapterkey{cxyz}  
\def\sectionkey{sxyz}  

\def\chapnum{\the\chapternumber}
\def\secnum{\the\sectionnumber}
\def\eqnum{\the\equationnumber}
\def\stnum{\the\statementnumber}
\def\fgnum{\the\figurenumber}

\def\acsname#1{\expandafter\aftergroup\csname#1\endcsname}
\def\xcsname#1{\expandafter\zcsname\the#1.\aftergroup.}
\def\zcsname#1{\ifx#1.\let\next=\relax\else\expandafter
\aftergroup\csname#1\endcsname\let\next=\zcsname\fi\next}
\def\ncsname#1{\ifx#1.\let\next=\relax\else\expandafter
\aftergroup\csname*#1\endcsname\let\next=\zcsname\fi\next}

\def\scxnum{\xcsname\chapternumber\xsectionnumber}
\def\eqxnum{\scxnum\xcsname\equationnumber}
\def\stxnum{\scxnum\xcsname\statementnumber}
\def\fgxnum{\xcsname\figurenumber}
\def\xsectionnumber{\expandafter\acsname{\the\sectionnumber}\aftergroup.}
\def\pgnum{\the\pageno}
\def\setcsnum#1.#2.{\begingroup\aftergroup\global\aftergroup\chapternumber
\ncsname#1.\aftergroup\global\aftergroup\sectionnumber\ncsname#2.\endgroup}

\def\secmode#1.#2.{\ifbookmode#1.\fi#2}
\def\sekmode#1.#2.{\ifbookmode#1.\fi\iflongmode#2.\fi}
\def\eqmode#1.#2.#3.{\ifbookmode#1.\fi\iflongmode#2.\fi#3}
\def\stmode#1 #2.#3.#4.{#1~\ifbookmode#2.\fi\iflongmode#3.\fi#4}
\def\stnmode#1 #2.#3.#4.{\ifbookmode#2.\fi\iflongmode#3.\fi#4}
\def\fgmode#1.{#1}

\def\chaptertop{20}\def\chaptermid{15}\def\chapterbot{30}

\def\Chapter#1.#2.#3.#4.{\Chapt#1.#2.#3.#4.#4.}

\def\Chapt#1.#2.#3.#4.#5.{{\numcat\global\beginningdocumentfalse
\global\pageno=#3
\global\chapternumber=#2 \global\sectionnumber=0
\global\equationnumber=1 \global\statementnumber=1
\expandafter\xdef\csname c*#1\endcsname{#2}
\gdef\chapterkey{#1} \gdef\sectionkey{#1}%
\ifeqnwrite\wdef{c*#1}\fi
\msg{Chapter \csname c*#1\endcsname.  Key = \chapterkey}%
\ifcontents
\write7{\csname Chapp\endcsname #2. #4.#3.}\fi
\page
\vglue\chaptertop pt \center \baselineskip=20pt \bigbf Chapter\kern.4em\the\chapternumber\par
\vglue\chaptermid pt \Bigbf\ignorespaces#5\par
\vglue\chapterbot pt}\HeaderNumber=0 \LeftHead #4.\RightHead #4.\firstpar}

\def\chapter#1{Chapter~\chapternum{#1}}

\def\chapternum#1{{\numcat\expandafter\ifx\csname c*#1\endcsname\relax
  \warn{Chapter #1 not defined}{\bf #1}\relax\else \csname c*#1\endcsname\fi}}

\def\Section#1#2.{{\numcat\gdef\sectionkey{#1}\ifbeginningdocument
   \else\global\advance\sectionnumber by 1\fi \global\beginningdocumentfalse
\begingroup\aftergroup\xdef\aftergroup\nuxxx\aftergroup{
\scxnum\aftergroup}\endgroup
\expandafter\ifx\csname s*#1\endcsname\relax\else
\edef\eqxxx{\csname s*#1\endcsname}%
\ifx\eqxxx\nuxxx\else\warna{Section #1 already exists}\fi\fi
\iflongmode\global\equationnumber=1\global\statementnumber=1\fi
\expandafter\xdef\csname s*#1\endcsname{\nuxxx} \ifeqnwrite\wdef{s*#1}\fi  
\msg{Section \sectionkey}\headtagleft{\expandafter\secmode\nuxxx. #2.}{#1}}\firstpar}

\def\section#1{Section~\sectionnum{#1}}

\def\sectionnum#1{{\numcat\expandafter\ifx\csname s*#1\endcsname\relax
  \warn{Section #1 not defined}{\bf #1}\else
  \edef\nuxxx{\csname s*#1\endcsname}\expandafter\secmode\nuxxx\fi}}

\def\Subsection#1.{\def\Subsecthead{#1}\Subhead{#1}\ifcontents
\write7{\csname Subsec\endcsname #1.\the\pageno.}\fi\Rightheadset\firstpar}

\def\Subsect#1.{\def\Subsecthead{#1}\Subhead{#1}\ifcontents
\write7{\csname Subsec\endcsname #1.\the\pageno.}\fi\firstpar}

\def\Rightheadset{\expandafter\RightHead \Subsecthead.}

\def\Appendix#1#2.{{\numcat\gdef\sectionkey{#1}\ifappendixmode
\global\advance\sectionnumber by 1\else
\global\sectionnumber = 91\global\appendixmodetrue\fi
\begingroup\aftergroup\xdef\aftergroup\nuxxx\aftergroup{
\scxnum\aftergroup}\endgroup
\expandafter\ifx\csname s*#1\endcsname\relax\else
\edef\eqxxx{\csname s*#1\endcsname}%
\ifx\eqxxx\nuxxx\else\warna{Section #1 already exists}\fi\fi
\iflongmode\global\equationnumber=1\global\statementnumber=1\fi
\expandafter\xdef\csname s*#1\endcsname{\nuxxx} \ifeqnwrite\wdef{s*#1}\fi  
\msg{Appendix \sectionkey}\headtagleft{Appendix \expandafter
\secmode\nuxxx. #2.}{#1}}\firstpar}

\def\appendix#1{Appendix~\appendixnum{#1}}

\def\appendixnum#1{{\numcat \expandafter\ifx\csname s*#1\endcsname\relax
  \warn{Section #1 not defined}{\bf #1}\else
  \edef\nuxxx{\csname s*#1\endcsname}\expandafter\secmode\nuxxx\fi}}

\def\addkeyz#1.#2\addkeyz{\ifx.#2.\sectionkey.#1.\else#1.#2\fi}

%
%
%

\def\Eq{\eqtag}
\def\eqtag#1{\newequation\eqno{#1}()}

\def\Eql#1#2$${\Eq{#1}\,\kern -\displaywidth
     \rlap{\quad $\displaystyle#2$}\kern \displaywidth \kern -1pt$$}
\def\El#1$${\eqno\,\kern -\displaywidth
     \rlap{\quad $\displaystyle#1$}\kern \displaywidth \kern -1pt$$}

\def\zlap#1{\llap{#1\kern-3pt}}

\def\newequation#1#2#3#4{\prooflabel{#2}{\numcat
\begingroup\aftergroup\xdef\aftergroup\nuxxx\aftergroup{
\eqxnum\aftergroup}\endgroup
\expandafter\ifx\csname e*#2\endcsname\relax\else
\edef\eqxxx{\csname e*#2\endcsname}%
\ifx\eqxxx\nuxxx\else\warna{Equation #2 already exists}\fi\fi
\expandafter\xdef\csname e*#2\endcsname{\nuxxx} \ifeqnwrite\wdef{e*#2}\fi}%
#1{#3\expandafter\eqmode\nuxxx#4}\global\advance\equationnumber by 1}

\def\eq#1{\eqcite{#1}()}

\def\eqcite#1#2#3{{\numcat\expandafter\ifx\csname e*#1\endcsname\relax
\warn{Equation #1 not defined.}\edef\eqxxx{#2#1\blackbox#3}\else
\edef\nuxxx{\csname e*#1\endcsname}%
\edef\eqxxx{#2\expandafter\eqmode\nuxxx#3}\fi{\rm\eqxxx}}}

\def\eqe{equation~\eq} \def\eqE{Equation~\eq} 
\def\eqs#1#2{\eq{#1},~\eq{#2}}

 \def\eqF{Formula~\eq}

\def\cth#1#2{#2{}\cr&\hskip#1pt{}#2}
  \def\ctq{\cth{20}} 
\def\creq{\ctq}

\def\addtab#1={#1\;&=}
\def\addtabe#1=#2={#1=#2\;&=}

\def\caeq#1#2{\def\\{\cr}\vcenter{\openup1\jot \halign
    {$\hfil\displaystyle ##\hfil$&&$\hfil\hskip#1pt\displaystyle ##\hfil$\cr#2\cr}}}
\def\ceq{\caeq{20}}

\def\ezeq#1#2#3{\def\\{\cr#1}\vcenter{\openup1\jot \halign{$\displaystyle 
   \hfil##$&$\displaystyle##\hfil$&&\hskip#2pt$\displaystyle##\hfil$\cr#1#3\cr}}}
\def\eaeq{\ezeq\addtab}

\def\eeq{\eaeq{20}}

\def\saeq#1#2{\def\\{\cr}\vcenter{\openup1\jot \halign{$\displaystyle
   ##\hfil$&&\hskip#1pt$\displaystyle##\hfil$\cr #2\cr}}}
\def\seq{\saeq{20}}

\def\qaeq#1#2{\def\\{&}\vcenter{\openup1\jot \halign{$\displaystyle
   ##\hfil$&&\hskip#1pt$\displaystyle##\hfil$\cr #2\cr}}}
\def\qeq{\qaeq{20}}

\def\ntable#1{\itemnumber=1 
   \def\\{\cr \the\itemnumber.\global\advance\itemnumber by 1 &}\vcenter{\openup1\jot 
   \halign{##&&\quad $\displaystyle{}##\hfil$\cr 
   \the\itemnumber.\global\advance\itemnumber by 1 &#1\cr}}}

\newdimen\itemindent \itemindent = 35pt
\def\hangitem{\hangindent\itemindent}
\def\itemo{\par\hangitem\textindent}
\def\items#1{\itemnumber=1 
   \def\\{\par \itemo{\ro(\the\itemnumber\ro)}\enspace\advance\itemnumber by 1}\\#1\par}
\def\ritems#1{\itemnumber=1 \def\\{\par \itemo{\ro (\it  
   \romannumber{\the\itemnumber}\/\ro )}\enspace\advance\itemnumber by 1}\\#1\par}
\def\aitems#1{\itemnumber=1  \def\\{\par \itemo{\ro (\sl 
  \alphanumber{\the\itemnumber}\/\ro )}\enspace\advance\itemnumber by 1}\\#1\par}
\def\mcases#1{\left\{\enspace\seq{#1}\right.}

%
%
%
%

\def\declare#1#2{\medbreak\indent{\bf#1.\kern.2em}#2\par\medbreak}
\def\exdeclare#1#2#3\par{\declare{\Statement{#1}{#2}}{#3}}
\def\stdeclare#1#2#3\par{\exdeclare{#1}{#2}{{\sl #3}}\par}

\def\Statement#1#2{\prooflabel{#2}{\numcat
\begingroup\aftergroup\xdef\aftergroup\nuxxx\aftergroup{\stxnum\aftergroup}\endgroup 
\edef\stxxx{#1 \nuxxx}%
\expandafter\ifx\csname t*#2\endcsname\relax\else
\edef\eqxxx{\csname t*#2\endcsname}%
\ifx\stxxx\eqxxx\else\warna{Statement #2 already exists}\fi\fi
\expandafter\xdef\csname t*#2\endcsname{\stxxx}%
\ifeqnwrite\wdef{t*#2}\fi}%
#1\kern.4em\expandafter\eqmode\nuxxx\global\advance\statementnumber by 1}

\def\stwrite#1#2{{\numcat\expandafter\ifx\csname t*#2\endcsname\relax
\warn{#1 #2 not defined.}\edef\stxxx{#1~#2\blackbox}\else
\edef\nuxxx{\csname t*#2\endcsname}%
\edef\stxxx{\expandafter\stmode\nuxxx}\fi{\stxxx}}}
\def\stnwrite#1#2{{\numcat\expandafter\ifx\csname t*#2\endcsname\relax
\warn{#1 #2 not defined.}\edef\stxxx{#2\blackbox}\else
\edef\nuxxx{\csname t*#2\endcsname}%
\edef\stxxx{\expandafter\stnmode\nuxxx}\fi{\stxxx}}}

\def\Ex#1{\exdeclare{Example}{#1}} \def\ex{\stwrite{Example}}

\def\Th#1{\stdeclare{Theorem}{#1}} \def\th{\stwrite{Theorem}}

\def\Lm#1{\stdeclare{Lemma}{#1}} \def\lm{\stwrite{Lemma}}

\def\Pr#1{\stdeclare{Proposition}{#1}} \def\pr{\stwrite{Proposition}}

\def\Co#1{\stdeclare{Corollary}{#1}} \def\co{\stwrite{Corollary}}

\def\Df#1{\exdeclare{Definition}{#1}}

\def\Labl#1{{\it#1\/}:\enspace}
\def\Label#1{\medbreak\indent\Labl{#1}}
\def\Labelpar #1.#2\par{\Label{#1}#2\par\medbreak}

\def\Proof{\Label{Proof}}

\def\Remark #1\par{\Labelpar Remark.#1\par}
\def\Warning #1\par{\Labelpar Warning.#1\par}
\def\Ack #1\par{\Labelpar Acknowledgments.#1\par}
\def\Acknowledgment #1\par{\Labelpar Acknowledgments.#1\par}

\def\qed{\hfill\parfillskip=0pt {\it Q.E.D.}\par\parfillskip=0pt plus 1fil\medbreak}

%
%
%
%
%

\newcount\refnumber \refnumber=1
\newif\ifrefmode \newif\ifrefs
\newif\ifrefcheck \refchecktrue
 
\def\refsin{{\refmodefalse \input\jobname.ref }\refstrue}

\def\refstyle{\normalbaselines
\parskip=\refparskip \frenchspacing
\rightskip=0pt plus 4em \spaceskip = .3333em \xspaceskip = .5em
\pretolerance=10000 \tolerance=10000
\hyphenpenalty=10000 \exhyphenpenalty=10000
\interlinepenalty=10000 \raggedbottom}

\def\References{{\refmodetrue {\center{\bigbf References}\par\nobreak} \vskip 15pt
{\refstyle \input \jobname.ref }}}

\def\book#1;#2;#3\par{\ifrefmode\namez#1;{\it #2},#3.\par\fi}
\def\paper#1;#2;#3; #4(#5)#6\par{\ifrefmode\namez#1;#2,{\it #3}\if a#4, to appear.\else
     { \bf #4}(#5),#6.\fi\par\fi}
\def\inbook#1;#2;#3;#4\par{\ifrefmode\namez#1;#2, {\sl in\/}:{\it #3},#4.\par\fi}
\def\appx#1;#2;#3;#4\par{\ifrefmode\namez#1;#2, {\sl appendix in\/}:{\it #3},#4.\par \fi}
\def\thesis#1;#2;#3\par{\ifrefmode\namez#1;{\it #2}, Ph.D.~Thesis,#3.\par\fi}
\def\uthesis#1;#2;#3\par{\ifrefmode\namez#1;{\it #2}, Ph.D.~Thesis, University of Minnesota,#3.\par\fi}
\def\preprint#1;#2;#3\par{\ifrefmode\namez#1;#2, preprint,#3.\par\fi}
\def\upreprint#1;#2;#3\par{\ifrefmode\namez#1;#2, preprint, University of Minnesota,#3.\par\fi}
\def\prepare#1;#2;#3\par{\ifrefmode\namez#1;#2, in preparation.\par\fi}
\def\personal#1;#2\par{\ifrefmode\namez#1; personal communication,#2.\par\fi}
\def\submit#1;#2;#3\par{\ifrefmode\namez#1;#2, submitted.\par\fi}
\def\other#1\par{\ifrefmode#1.\par\fi}

\def\key#1 {\ifrefmode\ifrefcheck
{\numcat\expandafter\ifx\csname w*#1\endcsname\relax\warna{Reference #1 not cited in text.}\else\fi}\fi
\hangindent2\parindent \textindent
{\ifrefs\ifproofmode{\fiverm #1} \fi\fi \rf{#1}}\else
{\numcat\expandafter\xdef\csname r*#1\endcsname{\the\refnumber}%
\global\advance\refnumber by 1\ifeqnwrite\wdef{r*#1}\fi}\fi}

\newif\ifnames \namesfalse

\def\namez#1;{\namesfalse\namezz#1,;} 
\def\namezz#1,#2,#3;{\ifx,#3,\ifnames \ and\fi\fi #1,#2,\ifx,#3,\else 
    \namestrue\namezz #3;\fi}

\def\namew#1;{\expandafter\nameww#1,,;} 
\def\nameww#1,#2,#3#4;{\write5{"#1,#2",\space\rnumberq\pageno}%
\ifx,#3\else\nameww#3#4;\fi}

\newskip\refspaceskip \refspaceskip=.25em
\newskip\refbrakskip \refbrakskip=.5pt
\newskip\refcommaskip \refcommaskip=2pt
\newskip\refminskip \refminskip=1pt

\def\rf#1{{\rm[\hglue\refbrakskip\rfn{#1}\hglue\refbrakskip]}}
\def\rfn#1{\rfvvv#1,\rfvvv}
\def\rfvvv#1,#2\rfvvv{\rfzzz#1;\rfzzz\ifx,#2,\else ,\hglue\refcommaskip\rfvvv#2\rfvvv\fi}
\def\rfzzz#1;#2\rfzzz{\def\rfyyy{\eatspace#1}%
{\sfcode`\.=1000\sfcode`\;=1000\sfcode`\:=1000\sfcode`\,=1000%
\spaceskip=\refspaceskip\numcat\edef\rfxxx{\rfyyy}\ifrefs\expandafter\ifx
\csname r*\rfyyy\endcsname\relax\warn{Reference  [\rfyyy]  not provided.}\else
\ifindex\expandafter\ifx\csname a*\eatspace#1\endcsname\relax
\warn{Authors for reference  [\rfyyy]  not provided.}\else
\expandafter\namew\csname a*\eatspace#1\endcsname;\fi\fi
\xdef\rfxxx{\csname r*\rfyyy\endcsname}\expandafter\xdef\csname
w*\rfyyy\endcsname{x}\ifrefmode\else
\ifrefnum\immediate\write3{[\rfxxx]}\fi\fi\fi\fi{\bf \rfxxx}\ifx;#2;\else;\nosemiz#2\fi}}

\def\rfr#1{[\hglue\refbrakskip\rfrn{#1}\hglue\refbrakskip]}
\def\rfrn#1{\rfrrr#1-}
\def\rfrrr#1-#2-{\rfn{#1}\hglue\refminskip--\hglue\refminskip\rfn{#2}}

\def\nosemiz#1;{#1}

\newif\ifrefnum \refnumfalse
\def\refnum{\immediate\openout3=\jobname.rfn\refnumtrue}

\newread\epsffilein    
\newif\ifepsffileok    
\newif\ifepsfbbfound   
\newif\ifepsfverbose   
\newdimen\epsfxsize    
\newdimen\epsfysize    
\newdimen\epsftsize    
\newdimen\epsfrsize    
\newdimen\epsftmp      
\newdimen\pspoints     
\pspoints=1bp          
\epsfxsize=0pt         
\epsfysize=0pt         
\def\texturesepsf{\def\specialstuff##1{illustration ##1 scaled \number\epsfscale}}

\def\epsfbox#1{\global\def\epsfllx{72}\global\def\epsflly{72}%
   \global\def\epsfurx{540}\global\def\epsfury{720}%
   \def\lbracket{[}\def\testit{#1}\ifx\testit\lbracket
   \let\next=\epsfgetlitbb\else\let\next=\epsfnormal\fi\next{#1}}%
\def\epsfgetlitbb#1#2 #3 #4 #5]#6{\epsfgrab #2 #3 #4 #5 .\\%
   \epsfsetgraph{#6}}%
\def\epsfnormal#1{\epsfgetbb{#1}\epsfsetgraph{#1}}%
\def\epsfgetbb#1{%
%
%
%
%
   {\epsffileoktrue \chardef\other=12
    \def\do##1{\catcode`##1=\other}\dospecials \catcode`\ =10
    \loop
       \read\epsffilein to \epsffileline
       \ifeof\epsffilein\epsffileokfalse\else
%
%
          \expandafter\epsfaux\epsffileline:. \\
        \fi
   \ifepsffileok\repeat
   \ifepsfbbfound\else
    \ifepsfverbose\message{No bounding box comment in #1; using defaults}\fi\fi}}%
%
%
\def\epsfsetgraph#1{%
   \epsfrsize=\epsfury\pspoints
   \advance\epsfrsize by-\epsflly\pspoints
   \epsftsize=\epsfurx\pspoints
   \advance\epsftsize by-\epsfllx\pspoints
%
%
   \epsfsize\epsftsize\epsfrsize
   \ifnum\epsfxsize=0 \ifnum\epsfysize=0
      \epsfxsize=\epsftsize \epsfysize=\epsfrsize
%
%
     \else\epsftmp=\epsftsize \divide\epsftmp\epsfrsize
       \epsfxsize=\epsfysize \multiply\epsfxsize\epsftmp
       \multiply\epsftmp\epsfrsize \advance\epsftsize-\epsftmp
       \epsftmp=\epsfysize
       \loop \advance\epsftsize\epsftsize \divide\epsftmp 2
       \ifnum\epsftmp>0
          \ifnum\epsftsize<\epsfrsize\else
             \advance\epsftsize-\epsfrsize \advance\epsfxsize\epsftmp \fi
       \repeat
     \fi
   \else\epsftmp=\epsfrsize \divide\epsftmp\epsftsize
     \epsfysize=\epsfxsize \multiply\epsfysize\epsftmp   
     \multiply\epsftmp\epsftsize \advance\epsfrsize-\epsftmp
     \epsftmp=\epsfxsize
     \loop \advance\epsfrsize\epsfrsize \divide\epsftmp 2
     \ifnum\epsftmp>0
        \ifnum\epsfrsize<\epsftsize\else
           \advance\epsfrsize-\epsftsize \advance\epsfysize\epsftmp \fi
     \repeat     
   \fi
%
%
   \ifepsfverbose\message{#1: width=\the\epsfxsize, height=\the\epsfysize}\fi
   \epsftmp=10\epsfxsize \divide\epsftmp\pspoints
   \vbox to\epsfysize{\vfil\hbox to\epsfxsize{%
    \special{\specialstuff{#1}}\hfil}}}%
%
%
{\catcode`\%=12 \global\let\epsfpercent=
%
%
\long\def\epsfaux#1#2:#3\\{\ifx#1\epsfpercent
   \def\testit{#2}\ifx\testit\epsfbblit
      \epsfgrab #3 . . . \\%
      \epsffileokfalse
      \global\epsfbbfoundtrue
   \fi\else\ifx#1\par\else\epsffileokfalse\fi\fi}%
%
%
\def\epsfgrab #1 #2 #3 #4 #5\\{%
   \global\def\epsfllx{#1}\ifx\epsfllx\empty
      \epsfgrab #2 #3 #4 #5 .\\\else
   \global\def\epsflly{#2}%
   \global\def\epsfurx{#3}\global\def\epsfury{#4}\fi}%
%
%

\newcount\epsfscale    
\newdimen\epsftmpp     
\newdimen\epsftmppp    
\newdimen\epsfM        
\newdimen\sppoints     
%
\epsfscale = 833  
\sppoints=1000sp       
\epsfM=1000\sppoints
%
\def\computescale#1#2{%
  \epsftmpp=#1 \epsftmppp=#2
  \epsftmp=\epsftmpp \divide\epsftmp\epsftmppp  
  \epsfscale=\epsfM \multiply\epsfscale\epsftmp 
  \multiply\epsftmp\epsftmppp                   
  \advance\epsftmpp-\epsftmp                    
  \epsftmp=\epsfM                               
  \loop \advance\epsftmpp\epsftmpp              
    \divide\epsftmp 2                           
    \ifnum\epsftmp>0
      \ifnum\epsftmpp<\epsftmppp\else           
        \advance\epsftmpp-\epsftmppp            
        \advance\epsfscale\epsftmp \fi          
  \repeat
  \divide\epsfscale\sppoints}
\def\epsfsize#1#2{%
  \ifnum\epsfscale=1000
    \ifnum\epsfxsize=0
      \ifnum\epsfysize=0
      \else \computescale{\epsfysize}{#2}
      \fi
    \else \computescale{\epsfxsize}{#1}
    \fi
  \else
    \epsfxsize=#1
    \divide\epsfxsize by 1000 \multiply\epsfxsize by \epsfscale
  \fi}


\newif\iffigs \figstrue
\def\figtopskip{10}\def\figbotskip{10}\def\figlabelskip{5}

\def\Fgg#1#2#3#4#5{\Figg{#1}{#2}{#3}{#4}{#5}Figure \expandafter\fgmode\nuxxx}

\def\Figg#1#2#3#4#5{\iffigs\openin\epsffilein=#1.ps \ifeof\epsffilein
\warn{Postscript file #1.ps not found}
\edef\nuxxx{#1\blackbox.}\else{\Figdraw{#1}{#2}{#3}{#4}{#5}}%
\closein\epsffilein\fi\else\edef\nuxxx{#1\blackbox.}\fi}

\def\Figdraw#1#2#3#4#5{{\numcat
\begingroup\aftergroup\xdef\aftergroup\nuxxx\aftergroup{\fgxnum\aftergroup}\endgroup 
\expandafter\ifx\csname f*#1\endcsname\relax\else
\edef\eqxxx{\csname f*#1\endcsname}%
\ifx\nuxxx\eqxxx\else\warna{Figure #1 already exists}\fi\fi
\expandafter\xdef\csname f*#1\endcsname{\nuxxx}\ifeqnwrite\wdef{f*#1}\fi}%
{\insert\topins{\penalty100 \splittopskip=0pt \splitmaxdepth=\maxdimen \floatingpenalty=0
\vglue #3pt\hfil \epsfbox{#1.ps}\hfill \par\nobreak \vglue#4pt%
\prooflabel{#1}\center{{\bf Figure \expandafter\fgmode\nuxxx.} \quad #2.}\par
\vglue#5pt}\global\advance\figurenumber by 1}}

\def\fgn#1{{\numcat\expandafter\ifx\csname f*#1\endcsname\relax
\warn{Figure #1 not defined.}\edef\nuxxx{#1\blackbox.}\else
\edef\nuxxx{\csname f*#1\endcsname}\fi \expandafter\fgmode\nuxxx}}

\def\Figbox#1{\openin\epsffilein=#1.ps \ifeof\epsffilein
\warn{Postscript file #1.ps not found}
\else  \hfil \epsfbox{#1.ps}\hfil
\closein\epsffilein\fi}  

\def\Figii#1#2#3#4#5{\Figdrawii{#1}{#2}{#3}{#4}{#5}Figure \expandafter\fgmode\nuxxx}

\def\Figdrawii#1#2#3#4#5{{\numcat
\begingroup\aftergroup\xdef\aftergroup\nuxxx\aftergroup{\fgxnum\aftergroup}\endgroup 
\expandafter\ifx\csname f*#1\endcsname\relax\else
\edef\eqxxx{\csname f*#1\endcsname}%
\ifx\nuxxx\eqxxx\else\warna{Figure #1 already exists}\fi\fi
\expandafter\xdef\csname f*#1\endcsname{\nuxxx}\ifeqnwrite\wdef{f*#1}\fi}%
{\insert\topins{\penalty100 \splittopskip=0pt \splitmaxdepth=\maxdimen \floatingpenalty=0
\vglue\figtopskip pt%
\epsfscale = #5
\Figbox{#3} \Figbox{#4} \par\nobreak
\vglue\figbotskip pt%
\prooflabel{#1}\center\center{{\bf Figure \expandafter\fgmode\nuxxx.} \quad#2.}\par
\vglue\figlabelskip pt}\global\advance\figurenumber by 1}}

\def\Figiv#1#2#3#4#5#6#7{{\numcat
\begingroup\aftergroup\xdef\aftergroup\nuxxx\aftergroup{\fgxnum\aftergroup}\endgroup 
\expandafter\ifx\csname f*#1\endcsname\relax\else
\edef\eqxxx{\csname f*#1\endcsname}%
\ifx\nuxxx\eqxxx\else\warna{Figure #1 already exists}\fi\fi
\expandafter\xdef\csname f*#1\endcsname{\nuxxx}\ifeqnwrite\wdef{f*#1}\fi}%
{\insert\topins{\penalty100 \splittopskip=0pt \splitmaxdepth=\maxdimen \floatingpenalty=0
\epsfscale = #7
\vglue\figtopskip pt%
\Figbox{#3} \Figbox{#4} \par\nobreak
\vglue\figbotskip pt\Figbox{#5} \Figbox{#6} \par\nobreak \vglue\figbotskip pt%
\prooflabel{#1}\center\center{{\bf Figure \expandafter\fgmode\nuxxx.} \quad#2.}\par
\vglue\figlabelskip pt}\global\advance\figurenumber by 1}Figure
\expandafter\fgmode\nuxxx}

%
%
%
%
%

\newif\ifindex \indexfalse
\newif\ifindexproof \indexprooffalse
\newif\ifcontents \contentsfalse

\def\contentsout{\contentstrue\immediate\openout7=\jobname.con}

\def\indexout{\indextrue
\immediate\openout4=\jobname.ind%
\openout5=\jobname.aut%
\openout6=\jobname.sym%
\numcat\input ref.aut\unnumcat}

\def\indexmargin#1{\ifindexproof\margintagleft{#1}\fi}

\def\ix#1{\ifindex\write4{"#1",\space\rnumberq\pageno}\fi\indexmargin{#1}}

\def\is#1{\iz{#1}\ix{#1}}



\fontdimen16\tensy=2.7pt \fontdimen17\tensy=2.7pt 

\def\sup#1{\raise 2.7pt \hbox{\sevenrm #1}}
\def\sub#1{\lower 2.7pt \hbox{\sevenrm #1}}

%
%
%

\def\ro#1{{\rm #1}} \def\rbox#1{\hbox{\rm #1}}
\def\roh{\rbox}

\def\roq#1{\qquad \rbox{#1}\qquad }

\def\iz#1{\expandafter\ifx\the\fam0{\it #1\/}\else {\rm #1}\fi}

\def\bo#1{{\bf #1}}

\font\bigrm=cmr12
\font\Bigrm=cmr10 scaled\magstep2
\font\bigit=cmmi12

\font\bigbf=cmbx12
\font\Bigbf=cmbx10 scaled\magstep2



\newfam\ibfam
\font\tenib=cmmib10 \textfont\ibfam=\tenib  
\font\sevenib=cmmib7 \scriptfont\ibfam=\sevenib
\font\fiveib=cmmib5 \scriptscriptfont\ibfam=\fiveib

\edef\ibhx{\hexnumber\ibfam}

\mathchardef\alphaB="0\ibhx0B
\mathchardef\betaB="0\ibhx0C
\mathchardef\gammaB="0\ibhx0D
\mathchardef\deltaB="0\ibhx0E
\mathchardef\epsilonB="0\ibhx0F
\mathchardef\zetaB="0\ibhx10
\mathchardef\etaB="0\ibhx11
\mathchardef\thetaB="0\ibhx12
\mathchardef\iotaB="0\ibhx13
\mathchardef\kappaB="0\ibhx14
\mathchardef\lambdaB="0\ibhx15
\mathchardef\muB="0\ibhx16
\mathchardef\nuB="0\ibhx17
\mathchardef\xiB="0\ibhx18
\mathchardef\piB="0\ibhx19
\mathchardef\rhoB="0\ibhx1A
\mathchardef\sigmaB="0\ibhx1B
\mathchardef\tauB="0\ibhx1C
\mathchardef\upsilonB="0\ibhx1D
\mathchardef\phiB="0\ibhx1E
\mathchardef\chiB="0\ibhx1F
\mathchardef\psiB="0\ibhx20
\mathchardef\omegaB="0\ibhx21
\mathchardef\varepsilonB="0\ibhx22
\mathchardef\varthetaB="0\ibhx23
\mathchardef\varpiB="0\ibhx24
\mathchardef\varrhoB="0\ibhx25
\mathchardef\varsigmaB="0\ibhx26
\mathchardef\varphiB="0\ibhx27
\mathchardef\partialB="0\ibhx40

\mathchardef\GammaBI="0\ibhx00
\mathchardef\DeltaBI="0\ibhx01
\mathchardef\ThetaBI="0\ibhx02
\mathchardef\LambdaBI="0\ibhx03
\mathchardef\XiBI="0\ibhx04
\mathchardef\PiBI="0\ibhx05
\mathchardef\SigmaBI="0\ibhx06
\mathchardef\UpsilonBI="0\ibhx07
\mathchardef\PhiBI="0\ibhx08
\mathchardef\PsiBI="0\ibhx09
\mathchardef\OmegaBI="0\ibhx0A


\newfam\euffam
\font\teneuf=eufm10 \textfont\euffam=\teneuf 
\font\seveneuf=eufm7 \scriptfont\euffam=\seveneuf 

\def\frak#1{{\fam\euffam\relax#1}}

\def\a{\frak a} \def\g{\frak g}  


\def\Cal#1{{\cal#1}}

\def\CA{\Cal A}  \def\CC{\Cal C} \def\CD{\Cal D}
\def\CE{\Cal E} \def\CF{\Cal F} 
\def\CG{\Cal G}  
 \def\CK{\Cal K} 
\def\CM{\Cal M} \def\CN{\Cal N} \def\CO{\Cal O}
\def\CP{\Cal P} \def\CQ{\Cal Q} 
 \def\CT{\Cal T}


\def\Gbar{\kern.1em \overline{\kern-.1em G}{}}

\def\hats{
\def\ahat{\hat a{}}\def\bhat{\hat b{}}\def\chat{\hat c{}}
\def\dhat{\hat d{}}\def\ehat{\hat e{}}\def\fhat{\hat f{}}
\def\ghat{\hat g{}}\def\hhat{\hat h{}}
\def\ihat{\hat \imath{}}\def\jhat{\hat \jmath{}}
\def\khat{\hat k{}}\def\lhat{\hat l{}}
\def\mhat{\hat m{}}\def\nhat{\hat n{}}\def\ohat{\hat o{}}
\def\phat{\hat p{}}\def\qhat{\hat q{}}\def\rhat{\hat r{}}
\def\shat{\hat s{}}\def\that{\hat t{}}\def\uhat{\hat u{}}
\def\vhat{\hat v{}}\def\what{\hat w{}}
\def\xhat{\hat x{}}\def\yhat{\hat y{}}\def\zhat{\hat z{}}}

\def\tildes{
\def\atilde{\tilde a{}}\def\btilde{\tilde b{}}\def\ctilde{\tilde c{}}
\def\dtilde{\tilde d{}}\def\etilde{\tilde e{}}\def\ftilde{\tilde f{}}
\def\gtilde{\tilde g{}}\def\htilde{\tilde h{}}
\def\itilde{\tilde \imath{}}\def\jtilde{\tilde \jmath{}}
\def\ktilde{\tilde k{}}\def\ltilde{\tilde l{}}
\def\mtilde{\tilde m{}}\def\ntilde{\tilde n{}}\def\otilde{\tilde o{}}
\def\ptilde{\tilde p{}}\def\qtilde{\tilde q{}}\def\rtilde{\tilde r{}}
\def\stilde{\tilde s{}}\def\ttilde{\tilde t{}}\def\utilde{\tilde u{}}
\def\vtilde{\tilde v{}}\def\wtilde{\tilde w{}}
\def\xtilde{\tilde x{}}\def\ytilde{\tilde y{}}\def\ztilde{\tilde z{}}}

\def\bars{
\def\abar{\bar a{}}\def\bbar{\bar b{}}\def\cbar{\bar c{}}
\def\dbar{\bar d{}}\def\ebar{\bar e{}}\def\fbar{\bar f{}}
\def\gbar{\bar g{}}\def\hbar{\bar h{}}
\def\ibar{\bar \imath{}}\def\jbar{\bar \jmath{}}
\def\kbar{\bar k{}}\def\lbar{\bar l{}}
\def\mbar{\bar m{}}\def\nbar{\bar n{}}\def\obar{\bar o{}}
\def\pbar{\bar p{}}\def\qbar{\bar q{}}\def\rbar{\bar r{}}
\def\sbar{\bar s{}}\def\tbar{\bar t{}}\def\ubar{\bar u{}}
\def\vbar{\bar v{}}\def\wbar{\mkern2mu \overline{\mkern-2mu w\mkern-1mu }\mkern1mu{}}
\def\xbar{\bar x{}}\def\ybar{\bar y{}}\def\zbar{\bar z{}}}

\def\narrowaccents{\tildes\hats\bars}

\narrowaccents


\newfam\msbfam 
\font\tenmsb=msbm10 \textfont\msbfam=\tenmsb  
\font\sevenmsb=msbm7 \scriptfont\msbfam=\sevenmsb

\edef\msbhx{\hexnumber\msbfam}
\def\Bbb#1{{\fam\msbfam\relax#1}}

\def\R{\Bbb R}  \def\C{\Bbb C}
  \def\P{\Bbb P}
\def\Q{\Bbb Q}  \def\Z{\Bbb Z}

\def\Rx{\R\mkern1mu^}  \def\Cx{\C\mkern1mu^}

\def\Rn{\Rx n}  \def\Cn{\Cx n}  
\def\Rm{\Rx m}  \def\Cm{\Cx m}


\newfam\exsfam
\font\eightex=cmex8 \textfont\exsfam=\eightex 
\font\sixex=cmex7 scaled 857 \scriptfont\exsfam=\sixex 

\edef\exshx{\hexnumber\exsfam}


\mathchardef\hbarq="0\msbhx7E   
\mathchardef\semidirect="2\msbhx6E
\mathchardef\directsemi="2\msbhx6F
\mathchardef\emptyset="0\msbhx3F
\mathchardef\subsetneq="2\msbhx28
\mathchardef\dagmath="0279
\mathchardef\ddagmath="027A
\mathchardef\Smath="0278
\mathchardef\Pmath="027B

\mathchardef\odotsy="220C
\mathchardef\oplussy="2208
\mathchardef\bigvee="1\exshx57
\mathchardef\bigwedge="1\exshx56
\mathchardef\bigcap="1\exshx54
\mathchardef\bigcup="1\exshx53
\mathchardef\bigotimes="1\exshx4E
\mathchardef\bigoplus="1\exshx4C
\mathchardef\bigodot="1\exshx4A
\mathchardef\Bigotimes="1\exshx4F
\mathchardef\Bigoplus="1\exshx4D
\mathchardef\Bigodot="1\exshx4B
\mathchardef\Bigwedge="1\exshx5E

\def\odot{\raise 1pt \hbox{$\,\scriptscriptstyle\odotsy\,$}}
\def\oplus{\raise 1pt \hbox{$\,\scriptscriptstyle\oplussy\,$}}
\def\tensor{\raise 1pt \hbox{$\,\scriptscriptstyle\otimes\,$}}
 
\def\Wedge{\raise 1pt \hbox{$\bigwedge$}}
\def\Odot{\raise 1pt \hbox{$\bigodot$}}
\def\Oplus{\raise 1pt \hbox{$\bigoplus$}}
\def\Tensor{\raise 1pt \hbox{$\bigotimes$}}

\def\comp{\raise 1pt \hbox{$\,\scriptstyle\circ\,$}}

\def\Upstar{^{\displaystyle *}}
\def\nequiv{\not\equiv}

\def\angles{\hbox{$\,<\kern-7.5pt\raise .7pt\hbox{$\scriptstyle)$}\,$}}

\def\Id{\font\elevenrm=cmr10 at 11pt\hbox{\tenrm 1\kern-3.8pt \elevenrm1}}

\def\simarrow{\setbox0 = \hbox{$\longrightarrow$} 
\setbox2 = \hbox to \wd0{\hfil$\widetilde {\hbox to .8 \wd0{\hfill}}\,$\hfil}
\wd2 = \wd0 \wd0=0pt \; \box0 \lower 1pt \box2\>}

\def\rddots{\mathinner{\mkern1mu\raise1pt\vbox{\kern7pt\hbox{.}}\mkern2mu
    \raise4pt\hbox{.}\mkern2mu\raise7pt\hbox{.}\mkern1mu}}

\def\blackbox{\vrule height7pt width5pt depth0pt}


\def\smallstyle{\font\ninerm=cmr9\font\ninei=cmmi9%
\font\ninett=cmtt9\font\ninesl=cmsl9\font\nineit=cmti9%
\font\ninesy=cmsy9\font\ninebf=cmbx9\font\ninesmc=cmcsc9%
\font\nineeuf=eufm9\font\ninemsb=msbm9\font\sevenex=cmex7%
\textfont0=\ninerm \scriptfont0=\sevenrm \scriptscriptfont0=\fiverm
\textfont1=\ninei \scriptfont1=\seveni \scriptscriptfont1=\fivei
\textfont2=\ninesy \scriptfont2=\sevensy \scriptscriptfont2=\fivesy
\textfont3=\sevenex \scriptfont3=\sevenex \scriptscriptfont3=\sevenex
\def\rm{\fam0\ninerm}%
\def\it{\fam\itfam\nineit}\textfont\itfam=\nineit
\def\sl{\fam\slfam\ninesl}\textfont\slfam=\ninesl
\def\bf{\fam\bffam\ninebf}\textfont\bffam=\ninebf 
\scriptfont\bffam=\sevenbf\scriptscriptfont\bffam=\fivebf
\def\tt{\fam\ttfam\ninett}\textfont\ttfam=\ninett
\textfont\msbfam=\ninemsb
\textfont\euffam=\teneuf
\setbox\strutbox=\hbox{\vrule height 8pt depth 3pt width 0pt}%
\lineskip=0pt\normalbaselineskip=10pt\parskip=2pt\normalbaselines\ninerm}



\def\ie{i.e., }
\def\eg{e.g., }
\def\cf{cf.~} \def \crf{\cf \rf}

\def\upth{\sup{th} }  \def\upst{\sup{st} }  
\def\xth #1{$#1$\upth}  
 \def\xsts #1{$(#1)$\upst} 
\def\ith{\xth i}  \def\jth{\xth j}  \def\kth{\xth k}
  \def\nth{\xth n}  \def\rth{\xth r}

\def\f#1{{\textstyle {1\over #1}}}
\def\fr#1#2{{\textstyle {#1\over #2}}}
\def\frac#1#2{{#1\over #2}}
\def\fra#1{{1\over #1}}
\def\supfr#1#2{\raise 6pt \hbox{$\scriptstyle #1\above.1pt
     \raise .4pt\hbox{$\scriptstyle #2$}$}}
\def\supp#1{\raise 7pt \hbox{$\scriptstyle #1$}}

\def\rg#1#2{#1=1,\ldots,#2}

\def\subs #1#2{#1_1,\ldots,#1_{#2}}
\def\psubs #1#2{(#1_1,\ldots,#1_{#2})}
\def\bsubs #1#2{\{#1_1,\ldots,#1_{#2}\}}
\def\csubs #1#2{[#1_1,\ldots,#1_{#2}]}

\def\psups #1#2{(#1^1,\ldots,#1^{#2})}

\def\ophantom#1#2{\setbox0=\hbox{$#1#2$}\setbox2 = \null
                 \ht2 = \ht0 \dp2 = \dp0 \box2}
\def\odphantom#1{\ophantom\displaystyle{#1}}
\def\otphantom#1{\ophantom\textstyle{#1}}

\def\set#1#2{\mathchoice{\left \{#1\odphantom{#2}\;\right |
\left .\;#2\odphantom{#1}\right \}}{\{#1\otphantom{#2}\,|\,#2\otphantom{#1} \}}
{\{#1\otphantom{#2}\,|\,#2\otphantom{#1} \}}
{\{#1\otphantom{#2}\,|\,#2\otphantom{#1} \}}}

\def\array#1{\null\,\vcenter{\normalbaselines\mathsurround=0pt
    \ialign{$##$\hfil&&\quad$##$\hfil\crcr\mathstrut\crcr\noalign{\kern-\baselineskip}
      #1\crcr\mathstrut\crcr\noalign{\kern-\baselineskip}}}\,}
\def\Array#1{\null\,\vcenter{\normalbaselines\mathsurround=0pt
    \ialign{$##$\hfil&&\qquad$##$\hfil\crcr\mathstrut\crcr\noalign{\kern-\baselineskip}
      #1\crcr\mathstrut\crcr\noalign{\kern-\baselineskip}}}\,}

\def\nnorm#1{\left \|\,#1\,\right \|}


\def\od#1#2{\mathchoice{d#1 \over d#2}{d#1/d#2}{d#1/d#2}{d#1/d#2}}
\def\odw#1#2{\mathchoice{d^2#1 \over d#2^2}%
{d^2#1 /d#2^2}{d^2#1 /d#2^2}{d^2#1 /d#2^2}}

\def\pd#1#2{\mathchoice{\partial #1 \over \partial #2}%
{\partial #1/\partial #2}{\partial #1/\partial #2}{\partial #1/\partial #2}}

\def\pds#1#2#3{\mathchoice{\partial^{#3} #1 \over \partial #2^{#3}}%
{\partial^{#3} #1/\partial #2^{#3}}{\partial^{#3} #1/\partial #2^{#3}}
{\partial^{#3} #1/\partial #2^{#3}}}

\def\mstrut{\vphantom{1_x}}

\def\sqsub#1_#2{\sqrt {\smash{#1_{#2}}\mstrut}}
\def\cusub#1_#2{\root 3 \of {\smash{#1_{#2}}\mstrut}}

\def\fpsub#1_#2#3#4{{\displaystyle #1_{#2}^{#3/#4}}}


\def\operator#1{\expandafter\def\csname#1\endcsname{\mathop{\rm #1}\nolimits}}

\operator{sign} \operator{mod}
\operator{Re} \operator{Im}
\operator{order} \operator{Order} \operator{ord}
\operator{Degree}
\operator{wt} \operator{weight}
\operator{tr} \operator{rank}
\operator{im} \operator{dom} \operator{supp}  
\operator{Span}
\operator{Ad} \operator{ad}
\operator{Hom} \operator{Ker}
\operator{Div} \operator{Grad} \operator{Curl}
\operator{div} \operator{grad} \operator{curl}
\operator{vol} \operator{Vol}

\operator{sech} \operator{csch}
\operator{arcsinh} \operator{arccosh}


 \def\gl#1{\frak{gl}(#1)}

 \def\sL#1{\frak{sl}(#1)}

 \def\glnR{\gl{n,\R}}

\def\An{\ro{A}(n)}


\def\at#1{|_{#1}}

 \def\andq{\roq{and}} 
\def\where{\roq{where}} 
 
\def\forall{\roq{for all}}

\def\J#1{\ro J{}^{#1}} \def\Jn{\J n}

\def\ps#1{^{(#1)}}

\def\un{u\ps n}  
\def\xun{(x,\un)}

%
%
%
%
%

\newdimen\headerskip \newbox\headerbox
\newdimen\headskip \headskip=7.25pt
\newdimen\headerdim
\newcount\HeaderNumber \HeaderNumber=0

\def\pagenobox{\hbox{\rnumbert\pageno}}

\newtoks\LeftHeader
\def\LeftHead#1.{\LeftHeader{#1}}
\newtoks\RightHeader
\def\RightHead#1.{\RightHeader{#1}}

\def\MakeHeaderbox{\setbox0 = \pagenobox
                   \setbox2 = \hbox to \wd0{\hfill}%
\ifodd\pageno
   \setbox\headerbox=\hbox to \hsize{\box2\hfill\tenit\the\RightHeader\hfill\box0}%
\else
   \setbox\headerbox=\hbox to \hsize{\box0\hfill\tenit\the\LeftHeader\hfill\box2}%
\fi
\ifnum\HeaderNumber=0 \setbox\headerbox=\hbox{\hfil} \fi
 \headerskip=\headskip \advance\headerskip by -\ht\headerbox
 \vbox to \headerdim{\vskip\headerskip\noindent\unhbox\headerbox\vskip-\headskip\vfill}%
 \nointerlineskip}

\def\MakeFooterbox{\ifproofmode\baselineskip = 24pt 
\line{\hss \tenrm \jobname\quad\shortdate\hss}\fi}

\def\PageBody{\vbox to\vsize{\boxmaxdepth\maxdepth \pagecontents}}
\def\OutputPage{
 \shipout\vbox{\MakeHeaderbox\PageBody\MakeFooterbox}
 \advancepageno\global\fnotenumber=0
 \ifnum\outputpenalty>-1000000 \else \dosupereject\fi
 \global\HeaderNumber=1}

\def\eatspace#1{\ifx#1 \eatspace\else#1\fi}

\long\def\eatspacepar#1{\ifx#1\par\let\next=\eatspacepar\else
\ifx#1 \let\next=\eatspacepar\else\let\next=#1\fi\fi\next}

%
%
%


\texturesepsf     